\documentclass[a4paper,12pt]{article}
\usepackage{amsmath}
\usepackage{amssymb}         
\usepackage{amsfonts}
\usepackage{times}
\usepackage{amsthm} 
\usepackage{epsf} 
\renewcommand{\theequation}
{\thesection.\arabic{equation}}         
\newcommand{\secct}[1]{\section{#1}
\setcounter{equation}{0}}               

\newtheorem{theorem}{Theorem}[section]         
\newtheorem{lemma}[theorem]{Lemma}             
\newtheorem{corollary}[theorem]{Corollary}     
\theoremstyle{plain}
\newcommand{\us}{{\underline s}}
\newcommand{\ux}{{\underline x}}
\newcommand{\uq}{{\underline q}}
\newcommand{\ut}{{\underline t}}
\newcommand{\uy}{{\underline y}}

\newcommand{\uQ}{{\underline Q}}
\newcommand{\umu}{{\underline \mu}}

\newcommand{\id}{{1\!\!{\rm I}}}
\newcommand{\sumtwo}[2]{\sum_{\substack{#1 \\ #2}}}
\renewcommand{\id}{{\mathchoice {\rm 1\mskip-4mu l} {\rm 1\mskip-4mu l} {\rm
1\mskip-4.5mu l} {\rm 1\mskip-5mu l}}}

\renewcommand{\us}{{\boldsymbol s}}
\renewcommand{\ux}{{\boldsymbol x}}
\renewcommand{\uq}{{\boldsymbol q}}
\renewcommand{\ut}{{\boldsymbol t}}
\renewcommand{\uy}{{\boldsymbol y}}

\newcommand{\flow}{{\boldsymbol{\mathcal J}}}
\renewcommand{\uQ}{{\boldsymbol Q}}
\renewcommand{\umu}{{\boldsymbol \mu}}

\newcommand{\uk}{{\boldsymbol k}}
\newcommand{\ul}{{\boldsymbol l}}

\begin{document}
\bibliographystyle{plain}
\setcounter{page}{0}
\thispagestyle{empty}

\title{Dissipative Transport:\\
Thermal Contacts and Tunnelling Junctions}

\author{
J\"{u}rg Fr\"{o}hlich$^{\rm 1}$\footnote{juerg@itp.phys.ethz.ch}\ ,\ \ \ Marco
Merkli$^{\rm 1}$\footnote{merkli@itp.phys.ethz.ch, supported by an NSERC Postoctoral
Fellowship and by ETH-Zurich}\ , \ \ \  Daniel Ueltschi$^{\rm
2}$\footnote{ueltschi@math.ucdavis.edu}\\
\ \\
$^{\rm 1}$Theoretische Physik \\
 ETH-H\"{o}nggerberg \\ 
CH-8093 Z\"{u}rich, Switzerland
\vspace*{.2cm}
\\
$^{\rm 2}$
Department of Mathematics\\
 University of California\\
Davis, CA 95616, USA }
\date{\today}
\maketitle

\begin{abstract}
The general theory of simple transport processes between quantum mechanical reservoirs is reviewed
and extended. We focus on thermoelectric phenomena, involving exchange of energy and particles.
The theory is illustrated on the example of two reservoirs of free fermions
coupled through a local interaction. 
We construct a stationary state and determine energy and particle currents
with the help of a convergent perturbation series.

We explicitly calculate several interesting quantities to lowest order, such as the entropy
production, the resistance, and the heat conductivity. Convergence of the
perturbation series allows us to prove that they are {\it strictly positive}
under suitable smallness and regularity assumptions on the interaction between the reservoirs.
\end{abstract}
\thispagestyle{empty}
\newpage
\tableofcontents

\secct{Introduction} \label{sec-1}

\subsection{Description of the problems}

Simple transport processes, such as those observed near a spatially localized
thermal contact or tunnelling junction between two macroscopically extended
metals at different temperatures and chemical potentials, have been studied
experimentally and theoretically for a long time; see e.g. [Ma]. One is
interested, for example, in measuring or predicting energy and charge
transport through a contact between two metals, as well as the rate of entropy
production. The natural theoretical description of such processes is provided
by quantum theory, more precisely by non-equilibrium quantum statistical
mechanics. The results of experiments or theoretical calculations can,
however, often be expressed in the language of thermodynamics. In this paper
we attempt to study such transport processes in a mathematically precise way,
extending or complementing results in [DFG, EPR, JP1, JP2, Ru1, Ru2]. \\
\indent
In recent years, interest in transport processes has been driven by various
experimental and theoretical developments in mesoscopic physics and the 
discovery of rather unexpected phenomena. Among them we mention
dissipation-free transport in incompressible Hall fluids [La, TKNN, BES, ASS,
FST], or in ballistic quantum wires [vW, Bee, ACF, FP1]. In such systems,
``transport in thermal equilibrium'' and the quantization of conductances are
observed. Further interesting transport processes are electron tunnelling into
an edge of a Hall fluid [CPW, CWCPW, LS, LSH], and tunnelling processes
between two different quantum Hall edges through a constriction leading to
measurements of fractional electric charges of quasi-particles (see, e.g.,
[SGJE], and [FP2] for theoretical considerations). At present, these processes
are only partially understood theoretically. Other examples are Josephson
junctions and Andreev scattering [SR1, SR2], or energy transport in chains
(see, e.g., [Af] and references given there).\\
\indent
In this paper, the main emphasis is put, however, on conceptual aspects of the
theory of simple dissipative transport processes between two
quantum-mechanical reservoirs and on an illustration of the general theory in
a simple example, namely transport of energy and charge between two 
metals, described as non-interacting electron liquids, at different
temperatures and chemical potentials. Of particular interest to us are
connections between theoretical descriptions based on non-equilibrium quantum
statistical mechanics, on the one hand, and on thermodynamics, on the other
hand. Our quantum-mechanical description involves equilibrium and
non-equilibrium states of macroscopic reservoirs with many degrees of
freedom. We show that, on intermediate time scales, tunnelling processes can
be described in terms of non-equilibrium stationary states (NESS), examples of
which have recently been studied in [DFG, EPR, JP1, JP2, Ru1, Ru2, BLR]. Our
construction of non-equilibrium stationary states is based on methods of
algebraic scattering theory and is inspired by ideas in [He, Rob, BR, BM,
Ha]. Links between quantum statistical mechanics and thermodynamics are
constructed by providing precise definitions of energy and particle currents
and of entropy production and by deriving a suitable form of the first and
second law of thermodynamics. The general theory developed in Sect.~2 reviews
ideas and results scattered over numerous articles and books and represents an
attempt to provide a somewhat novel and, we believe, rather clear synthesis. A
more complete version, including the treatment of systems with time-dependent Hamiltonians, appears in [FMSU] and in a forthcoming paper. It is
illustrated on the example of two reservoirs of non-interacting electrons
coupled through local many-body interactions (Sects 3 through 5). Examples of
non-equilibrium stationary states supporting particle and/or energy currents
are constructed with the help of a convergent perturbation (Dyson) series in
the many-body interaction terms. The currents and the entropy production rate
are calculated quite explicitly to leading order. This enables us to show
that, under natural hypotheses, they are strictly positive. Onsager
reciprocity relations are established to lowest non-trivial order in the
many-body interaction terms. Positivity of the entropy production rate has
also been established recently in [AP, MO] for XY chains, and in [CNP] for
wave turbulence. 

\subsection{Contents of paper}

In Section 2.1, quantum-mechanical reservoirs are introduced, whose time
evolution is given in terms of a one-parameter group of $*$automorphisms of a
kinematical $C^*$-algebra of operators. Conservation laws of reservoirs are
described by commuting conserved charges. The equilibrium states of such
reservoirs are introduced and parametrized by temperature and chemical
potentials. Two general assumptions, (A1) and (A2), are formulated. They state
that the thermodynamic limits of the time evolution and of the gauge
transformations of operators in the kinematical algebra of an infinite system
and of the equilibrium states exist.\\
\indent
In Section 2.2 we study two interacting reservoirs at different temperatures
and chemical potentials. Each reservoir is required to satisfy the assumptions
formulated in Sect. 2.1. A class of many-body interactions coupling the two
reservoirs is introduced. It is assumed that the thermodynamic limit of the
interactions and of the corresponding time evolutions exist. Energy and charge
currents for finite and infinite systems are then defined.\\
\indent
Connections between quantum statistical mechanics and thermodynamics are
elucidated in Section 2.3. The entropy production rate is defined and
expressed in terms of the currents and of thermodynamic parameters. An
inequality expressing the positivity of relative entropy is shown to imply
that the total entropy production is non-negative (see also [Ru2, JP1]). \\
\indent
In Section 2.4, non-equilibrium stationary states for coupled reservoirs are
introduced. They can be construced with the help of scattering (M\o ller)
endomorphisms of the kinematical algebra of the infinite coupled system. A
precise assumption concerning the existence of scattering endomorphisms is
formulated. Our approach has its roots in Hepp's work on the Kondo problem
[He] and Robinson's analysis of return to equilibrium in the XY spin chain
[Rob]. Robinson's ideas have been put into a general context in [BR, Ha]. The
scattering approach is the starting point for numerous heuristic studies of
thermal contacts and tunnelling junctions (see, e.g., [Ma]). The first
mathematically rigorous implementation of this approach in a study of energy
transport in the XY spin chain and of tunnelling between free-fermion
reservoirs appeared in [DFG].\\
\indent
In Section 2.5, long-time stability properties of equilibrium and
non-equilibrium stationary states against perturbations of the initial state
of the coupled system are studied, and conditions for the existence of
temperature or density profiles in non-equilibrium stationary states are
identified. \\
\indent
The general theory of Section 2 is illustrated in Sections 3,4 and 5 on the
example of two coupled free-electron reservoirs. \\
\indent
In Section 3, the quantum theory of finite and infinite reservoirs of free
electrons is briefly recalled, and a class of local many-body interactions
between two such reservoirs satisfying the general assumptions formulated in
Section 2 is introduced.\\
\indent
Our main technical result, the existence of scattering (M\o ller)
endomorphisms, defined on an appropriate kinematical $C^*$-algebra describing
two infinite free-electron reservoirs, is established in Section 4. A similar
result has previously been proven in [BM]. We show that, under appropriate
smallness and regularity assumptions on the many-body interaction terms, the
scattering endomorphisms are given by a (norm-) convergent Dyson series. As a
consequence, non-equilibrium stationary states can be constructed with the
help of a convergent perturbation expansion.\\
\indent
The results of Section 4 are used in Section 5 to derive explicit expressions
for the energy and charged-particle currents to leading order in the many-body
interaction terms. These expresssions, along with the convergence of the Dyson
series, prove that, for small coupling constants, the entropy production rate
is strictly positive, Ohm's law holds to leading order in the voltage drop
between the reservoirs, with a resistance whose temperature dependence can be
determined, and the Onsager reciprocity relations hold to leading order. \\
\indent
We conclude this introduction with explicit formulae for the leading
contributions to the particle current, ${\cal J}$, and to the energy current,
${\cal P}$, between two reservoirs, $I$ and $II$, of free electrons coupled to
each other by a quadratic local interaction term with a form factor ${\widehat
  w}((-\uk, II), (\ul,I))$ and a brief discussion of the qualitative
implications of these formulae. These currents are given by
\begin{eqnarray*}
{\cal J}\!\!&\simeq&\!\!2\pi\!\!\int_{{\mathbb R}^6} d \uk\ d\ul\  \delta(|\uk|^2-|\ul|^2)
\left|\widehat{w}((-\uk,II),
  (\ul,I))\right|^2\left(\rho_{II}(\uk)-\rho_I(\uk)\right),\nonumber\\
{\cal P}\!\!&\simeq&\!\!2\pi\!\!\int_{{\mathbb R}^6} d\uk\ d\ul\
|\uk|^2\delta(|\uk|^2-|\ul|^2) 
\left|\widehat{w}((-\uk,II),
  (\ul,I))\right|^2\left(\rho_{II}(\uk)-\rho_I(\uk)\right),
\hfill(*)
\end{eqnarray*}
where $\rho_r(\uk)$ is the Fermi distribution of the free electron gas ($r=I,
II$ labels the reservoirs), and
$\widehat{w}((-\uk,II),(\ul,I))$ is the interaction kernel describing
scattering of a particle in an initial state with energy $|\ul|^2$ localized in
reservoir $I$ to a final state with energy $|\uk|^2$ localized in reservoir
$II$.\\ 
\indent
If both reservoirs have the same chemical potential and the temperatures
satisfy $T^I<T^{II}$ then ${\cal J}$ and ${\cal P}$ are positive;
particles and energy flow from the hotter to the colder reservoir. Similarly,
if the reservoirs have the same temperature, then particles and energy flow
from the reservoir with the higher chemical potential to the other one. \\
\indent
Formulae ($*$) prove that the leading contribution to
the entropy production rate is strictly positive, unless both reservoirs are
at the same temperature and at the same chemical potential.\\
\indent
Another consequence of ($*$) is that, to leading order in the
interaction, and for a small voltage drop,
$\Delta\mu=\mu^{II}-\mu^I$ (at a fixed temperature, $T$, for both
reservoirs), {\it Ohm's law} is valid, i.e., the voltage drop is proportional to
the current,  
\begin{equation*}
\Delta\mu\simeq R(\mu^I,T) {\cal J}.
\end{equation*}
Our calculations show that the resistance $R(\mu^I,T)$ grows linearly in $T$,
for large $T$, it has a 
positive value at $T=0$ and may increase or decrease in $T$, at small
temperatures, depending on properties of the interaction kernel
$\widehat{w}$ modelling the junction between the two reservoirs.

\vspace{3mm}

This paper is dedicated to {\it Klaus Hepp} and {\it David Ruelle} on the occasion of their retirement from active duty, but not from scientific activity. Some of their work plays a significant r\^ole in the analysis presented in this paper. J. F. is deeply grateful to them for their generous support and for everything they have contributed to making his professional life at ETH and at I.H.E.S. so pleasant.\\

{\it Acknowledgements.\ } J.F. thanks S. Dirren and G. M. Graf for useful discussions
during an early stage of his efforts. We are grateful to R. Fern\'andez, B. Nachtergaele and I. M. Sigal for numerous helpful discussions. 
We have greatly benefitted from studying the works of V. Jak\v si\'c and C.-A.
Pillet
and of D. Ruelle on related problems. D.U. acknowledges the hospitality of ETH
Z\"urich, where most of this work has been carried out.

\secct{Elements of a general theory of junctions and of
  non-equilibrium stationary states} \label{sec-2}

\subsection{Quantum theory of reservoirs} \label{subsec-2.1}

We start our general analysis by describing {\it
  ``quantum-mech\-an\-ic\-al reservoirs''}. A reservoir is a quantum system
with very many degrees of freedom, e.g., an electron liquid in a
normal metal, a superconductor, a gas of atoms, or a large array of
coupled, localized spins, but with a small number of observable
physical quantities. It is confined to a macroscopically 
large, but compact subset, $\Lambda$, of physical space ${\mathbb
  R}^3$. Its pure states correspond, as usual, to unit rays in a {\it
  separable Hilbert space}, ${\mathcal H}^\Lambda$, and its dynamics is
  generated by a selfadjoint {\it Hamiltonian}, $H^\Lambda$, acting on the
  space ${\mathcal H}^\Lambda$. The kinematics of the reservoir is encoded into an algebra, ${\mathcal A}^\Lambda$, 
  of operators contained in (or equal to) the algebra of all bounded
  operators on ${\mathcal H}^\Lambda$ \footnote{The algebra ${\mathcal A}^\Lambda$ is sometimes
  called algebra of ``observables'', a commonly used, but unfortunate expression.}.
  The time evolution of an
  operator $a$ on ${\mathcal H}^\Lambda$ is given, in the
  Heisenberg picture, by
\begin{equation}
\alpha_t^\Lambda (a)\;:=\; e^{it H^\Lambda/\hbar}
~a\,e^{-itH^\Lambda/\hbar}, 
\label{eq:2.1}
\end{equation}
and it is assumed that $\alpha_t^\Lambda (a) \in {\mathcal
  A}^\Lambda$, for every $a \in {\mathcal A}^\Lambda$.

There may exist a certain number of linearly independent, commuting
{\it conservation laws},
which are represented by selfadjoint operators $Q_1^\Lambda, \ldots,
Q_M^\Lambda$ on ${\mathcal H}^\Lambda$ commuting with the dynamics of
the reservoir, i.e., 
\begin{equation}
\left[ H^\Lambda, Q_j^\Lambda\right]\;=\; 0,~ 
\left[ Q_i^\Lambda, Q_j^\Lambda\right]\;=\; 0,~
\left[ Q_i^\Lambda, a\right]\;=\;0, 
\label{eq:2.2}
\end{equation}
for all $i,j,=1,\ldots,M$, and for all ``observables'' $a \in
{\mathcal A}^\Lambda$. More precisely, one assumes that all operators
$\exp i t H^\Lambda/\hbar, \left\{ \exp i\,
  s_j\,Q_j^\Lambda\right\}_{j=1}^M$ commute with one another, for
arbitrary real values of $t, s_1, \ldots, s_M$.

A typical example of a conservation law is the {\it particle number
  operator}, $N^\Lambda$, of a reservoir consisting of a gas of
non-relativistic atoms.

On the algebra, $B({\mathcal H}^\Lambda)$, of all bounded
operators on the Hilbert space ${\mathcal H}^\Lambda$, we define
{\it ``gauge transformations of the first kind''} by setting
\begin{equation}
\varphi_\us^\Lambda (a)\;:=\; e^{i\us\cdot
  \uQ^\Lambda}~ a\;e^{-i\,\us\cdot\uQ^\Lambda},
\label{eq:2.3}
\end{equation}
for $a \in B ({\mathcal H}^\Lambda)$, where
\begin{equation}
\us \cdot \uQ^\Lambda \;:=\; \sum_{j=1}^M s_j\;Q_j^\Lambda~.
\label{eq:2.4}
\end{equation}
Then $\left\{ \varphi_{\us}^\Lambda \bigm| \us \in {\mathbb
    R}^M\right\}$ is an $M$-parameter abelian group of
$^*$automorphisms (see (\ref{eq:2.15}), below) of the algebra
$B({\mathcal H}^\Lambda)$, and 
\begin{equation}
\alpha_t^\Lambda \left( \varphi_{\us}^\Lambda (a)\right) \;=\;
\varphi_\us^\Lambda \left( \alpha_t^\Lambda (a)\right), 
\label{eq:2.5} 
\end{equation}
for all $a\in B ({\mathcal H}^\Lambda)$, by (\ref{eq:2.2}). It is
natural to define the ``observable algebra'' ${\mathcal A}^\Lambda$ as
the algebra of all those operators $a \in B ({\mathcal H}^\Lambda)$
for which
\begin{equation}
\varphi_\us^\Lambda (a)\;=\;a\,, ~{\rm for~all~~} \us \in {\mathbb
  R}^M. 
\label{eq:2.6}
\end{equation}

To every conservation law $Q_j^\Lambda$ there corresponds a conjugate
thermodynamic parameter, $\mu_j$, commonly called a {\it chemical
  potential}. 

{\it Thermal equilibrium} of the reservoir at inverse temperature
$\beta$ and chemical potentials $\umu = (\mu_1,\ldots,\mu_M)$ is
described by a mixed state, or {\it density matrix}, given by
\begin{equation}
\left( \Xi_{\beta,\umu}^\Lambda\right)^{-1} \exp -\beta \left[
  H^\Lambda - \umu \cdot \uQ^\Lambda\right],
\label{eq:2.7}
\end{equation}
where
\begin{equation}
\Xi_{\beta,\umu}^\Lambda \;:=\;{\rm tr} \left( \exp -\beta \left[ H^\Lambda
    - \umu\cdot\uQ^\Lambda \right]\right)
\label{eq:2.8}
\end{equation}
is the grand-canonical partition function. Of course, it is assumed that the
operators 
\begin{equation*}
\exp -\beta \left[ H^\Lambda - \umu\cdot\uQ^\Lambda\right]
\end{equation*}
are trace-class, for all chemical potentials $\umu$ in a region
${\mathcal M} \subseteq {\mathbb R}^M$, for all $\beta>0$, and for
arbitrary compact subsets $\Lambda$ of physical space ${\mathbb
  R}^3$. It is also commonly assumed that reservoirs are {\it
  thermodynamically stable}, in the sense that the thermodynamic
potential, $G$, given by
\begin{equation}
\beta G \left( \beta, \umu, V\right) \;:=\; - \ln \left(
  \Xi_{\beta,\umu}^\Lambda\right) 
\label{eq:2.9}
\end{equation}
is {\it extensive}, i.e., proportional to the volume, $V$, of the set
$\Lambda$, up to boundary corrections, for arbitrary $\beta >0, \umu
\in {\mathcal M}$. 

The expectation value of an operator $a \in B ({\mathcal H}^\Lambda)$
in the equilibrium state corresponding to the density matrix
(\ref{eq:2.7}) is given by
\begin{equation}
\omega_{\beta,\umu}^\Lambda (a) \;:=\; \left(
  \Xi_{\beta,\umu}^\Lambda\right)^{-1} {\rm tr}\,\left(
  e^{-\beta\left[ H^\Lambda -\umu\cdot\uQ^\Lambda\right]}\,a\right).
\label{eq:2.10}
\end{equation}
The state $\omega_{\beta,\umu}^\Lambda$ has some remarkable properties
to be discussed next.

It is {\it time-translation invariant}, i.e.,
\begin{equation}
\omega_{\beta,\umu}^\Lambda \left(\alpha_t^\Lambda(a)\right)\;=\;
\omega_{\beta,\umu}^\Lambda (a),
\label{eq:2.11}
\end{equation}
for arbitrary $a\in B({\mathcal H}^\Lambda)$. It obeys the celebrated
KMS {\it condition}
\begin{equation}
\omega_{\beta,\umu}^\Lambda \left( \alpha_t^\Lambda (a)\, b\right) \;=\;
\omega_{\beta,\umu}^\Lambda \left( b\,\alpha_{t+i\beta\hbar}^\Lambda\,
  \left( \varphi_{-i\beta\umu}^\Lambda (a)\right)\right),
\label{eq:2.12}
\end{equation}
for arbitrary $a$ and $b$ in $B({\mathcal H}^\Lambda)$. In particular,
if $a\in{\mathcal A}^\Lambda$ then
\begin{equation}
\omega_{\beta,\umu}^\Lambda \left( \alpha_t^\Lambda(a)\,b\right) \;=\;
\omega_{\beta,\umu}^\Lambda \left( b\,\alpha_{t+i\beta\hbar}^\Lambda
  \, (a)\right),
\label{eq:2.13}
\end{equation}
for arbitrary $b\in B({\mathcal H}^\Lambda)$; see
(\ref{eq:2.6}). Eq.~(\ref{eq:2.12}) is an easy consequence of
eqs.~(\ref{eq:2.10}) and (\ref{eq:2.3}) and of the
cyclicity of the trace, i.e.,
\[
{\rm tr}\,(a\, b) \;=\; {\rm tr}\, (b\,a).
\]
For further details concerning these standard facts of quantum
statistical 
mechanics we refer the reader to [Ru 3, BR].

Next, we recall some conventional wisdom concerning the {\it
  thermodynamic limit} of a reservoir. We are interested in
understanding asymptotics of physical quantities, as the region
$\Lambda$ to which the reservoir is confined increases to all of
${\mathbb R}^3$, or to an infinite half-space
\begin{equation}
{\mathbb R}_\pm^3 \;:=\; \left\{\left( x,y,z\right) \in {\mathbb R}^3
  \bigm| x \gtrless 0\right\}.
\label{halfspace}
\end{equation}
We use the notation~ ``$\Lambda \nearrow \infty$''~ to mean that
$\Lambda \nearrow {\mathbb R}^3$ (or $\Lambda \nearrow {\mathbb
  R}_\pm^3$), in the sense of Fisher (meaning, in essence, that the ratio between the surface and the
  volume goes to zero); see [Ru 3].

We introduce an operator algebra ${\mathcal F}^r$, called the {\it
  ``field algebra''}, convenient for 
the description of the thermodynamic limit of a reservoir:
\begin{equation}
{\mathcal F}^r \;:=\; \overline{ \bigvee_{\Lambda
    \nearrow \infty} B({\mathcal H}^\Lambda)}, 
\label{eq:2.14}
\end{equation}
where $\bigvee_{\Lambda\nearrow\infty} B({\mathcal
  H}^\Lambda)$ is the algebra generated by all the operators in the
increasing sequence of algebras 
\[
\ldots \subseteq\; B \big({\mathcal H}^\Lambda \big) \;\subseteq\; B
\big({\mathcal H}^{\Lambda'}\big) \;\subseteq\; \ldots, 
\]
$\Lambda \subseteq \Lambda'$, and $\overline{(\cdot)}$
denotes the closure in the operator norm. Technically speaking,
${\mathcal F}^r$ is a C$^*$-algebra, [BR]. The superscript $^r$ stands for
``reservoir''. Below, we consider two interacting reservoirs, labelled
by $r=I,II$.

A group $\{ \tau_\ut \bigm| \ut \in {\mathbb R}^n\}$ of {\it
homomorphisms}, of a C$^*$-algebra ${\mathcal F}$ is an
{\it n-parameter $^*$automorphism group} of ${\mathcal F}$ iff
\[
\tau_{\ut=0} (a)\;=\;a,~~ \tau_\ut \left( \tau_{\ut'} (a)\right) \;=\;
\tau_{\ut + \ut'} (a),
\]
and
\begin{equation}
\tau_\ut (a)^* \;=\; \tau_\ut (a^*),
\label{eq:2.15}
\end{equation}
for all $a\in {\mathcal F}$ and arbitrary $\ut, \ut' \in {\mathbb
  R}^n$.
  
For the purposes of this paper we shall require the following two 
  assumptions concerning the existence of the thermodynamic limit.

\vspace{.5cm}

{\bf (A1)~ Existence of the thermodynamic limit of the dynamics and the gauge transformations}.

{\it
For every operator $a\in{\mathcal F}^r$, the limits in
operator norm
\begin{equation}
\displaystyle\mathop{n-\lim}_{\Lambda\nearrow\infty}\; \alpha_t^\Lambda
(a) \;=:\; \alpha_t (a)
\label{eq:2.16}
\end{equation}
and
\begin{equation}
\displaystyle\mathop{n-\lim}_{\Lambda\nearrow\infty}\;
\varphi_\us^\Lambda (a) \;=:\; \varphi_\us (a)
\label{eq:2.17}
\end{equation}
exist, for all $t\in{\mathbb R},~\us \in{\mathbb R}^M$, and define
$^*$automorphism groups of the field algebra ${\mathcal F}^r$. The
convergence in (\ref{eq:2.16}) and (\ref{eq:2.17}) is assumed to be
{\it uniform} for $t$ in any compact interval
of ${\mathbb R}$ and for $\us$ in any compact subset of ${\mathbb
  R}^M$, respectively. The $^*$automorphism groups $\alpha_t$ and
$\varphi_\us$ may be assumed to be norm-continuous in $t$ and $\us$,
respectively; (but ``weak$^*$ continuity'' will usually be
sufficient).
}\\

\noindent
We define the {\it ``kinematical algebra''}, ${\mathcal A}^r$, to be the
largest subalgebra of ${\mathcal F}^r$ {\it pointwise invariant} under
$\{\varphi_\us\}_{\us\in{\mathbb R}^M}$, i.e., 
\begin{equation}
{\mathcal A}^r\; :=\; \left\{ a\in{\mathcal F}^r\bigm| \varphi_\us
  (a)\;=\; a, ~{\rm for~ all~~} \us\in{\mathbb R}^M\right\}.
\label{eq:2.18}
\end{equation}
Since, by (\ref{eq:2.5}), (\ref{eq:2.16}) and (\ref{eq:2.17}),
$\alpha_t$ and $\varphi_\us$ commute, $\alpha_t (a) \in {\mathcal A}^r$,
for every $a\in{\mathcal A}^r$.

\vspace{.5cm}

{\bf (A2)~ Existence of the thermodynamic limit of the equilibrium state}.

{\it
For every $a\in{\mathcal F}^r$,
\begin{equation}
\lim_{\Lambda\nearrow\infty} \;\omega_{\beta,\umu}^\Lambda (a)\;=:\;
\omega_{\beta,\umu} (a)
\label{eq:2.19}
\end{equation}
exists and is time-translation invariant, i.e., 
\begin{equation}
\omega_{\beta,\umu} \,\left( \alpha_t (a)\right) \;=\;
\omega_{\beta,\umu} (a),
\label{eq:2.20}
\end{equation}
for all $a\in{\mathcal F}^r, ~t\in{\mathbb R}\,$.
}\\

\noindent
We assume that ${\mathcal F}^r$ contains a norm-dense subalgebra
$\overset{\;\circ}{{\mathcal F}^r}$ with the property that the operator
$\alpha_t (\varphi_\us (a))$ extends to an entire function of
$t\in{\mathbb C},~\us\in{\mathbb C}^M$, for every $a\in\overset{\;\circ}{{\mathcal F}^r}$. If
$\alpha_t$ and $\tau_\us$ are norm-continuous in $t$ and $\us$,
respectively, the existence of an algebra $\overset{\;\circ}{{\mathcal F}^r} \subset {\mathcal F}^r$
with these properties is an easy theorem.

{}From the KMS condition (\ref{eq:2.12}), and from (\ref{eq:2.16}),
(\ref{eq:2.17}), (\ref{eq:2.19}), it follows that the infinite-volume state
$\omega_{\beta,\umu}$ on ${\mathcal F}^r$ obeys the KMS {\it condition}
\begin{equation}
\omega_{\beta,\umu} (\alpha_t(a)\, b )\;=\; 
\omega_{\beta,\umu} ( b\;\alpha_{t+i\beta\hbar}\;( 
\varphi_{-i\beta\umu} (a))), 
\label{eq:2.21}
\end{equation}
for all $a\in\overset{\;\circ}{{\mathcal F}^r},~ b\in{\mathcal F}^r$. If  $a\in{\mathcal A}^r\cap\overset{\;\circ}{{\mathcal F}^r}$
then eq.~(\ref{eq:2.21}) simplifies to
\begin{equation}
\omega_{\beta,\umu} (\alpha_t(a)\,b)\;=\;
\omega_{\beta,\umu} ( b\;\alpha_{t+i\beta\hbar} (a)) .
\label{eq:2.22}
\end{equation}

\subsection{Thermal contacts and tunnelling junctions between
  macroscopic reservoirs} \label{subsec-2.2}

We consider two reservoirs, $I$ and $I\!I$, with all the
properties described in Sect.~2.1. These reservoirs may or may not
have the same physical properties. 
For example, they may be ordinary metals located in two complementary
half-spaces of ${\mathbb R}^3$; or $I$ may be a metal and $I\!I$ a
superconductor, etc. Later, we shall consider the example where $I$ and
$I\!I$ are ordinary metals, i.e., non-interacting electron liquids. In the
following, ``$I$'' will be a shorthand notation for $(I,\Lambda^I)$, and
``$I\!I$'' for $(I\!I,\Lambda^{I\!I})$, where $\Lambda^I$ and $\Lambda^{I\!I}$
are arbitrary compact subsets of ${\mathbb R}^3$. Realistically
$\Lambda^I$ and $\Lambda^{I\!I}$ should not intersect;
but we shall ignore this constraint.

The Hilbert space of the system obtained by composing the two
reservoirs is given by
\begin{equation}
{\mathcal H}\;=\;{\mathcal H}^I\otimes {\mathcal H}^{I\!I}
\label{eq:2.31}
\end{equation}
and the dynamics, {\it before} the reservoirs are brought into
contact, is generated by the Hamiltonian
\begin{equation}
H^0 \;:=\; H^I \otimes \id\;+\;\id\otimes H^{I\!I}. 
\label{eq:2.32}
\end{equation}
The natural algebra of operators of the coupled system is given by
$B({\mathcal H}^I) \otimes B({\mathcal H}^{I\!I})$ and, in the
thermodynamic limit ($\Lambda^I      \nearrow{\mathbb R}_-^3$ and
$\Lambda^{I\!I} \nearrow{\mathbb R}_+^3$, or
$\Lambda^I      \nearrow{\mathbb R}^3$ and 
$\Lambda^{I\!I} \nearrow{\mathbb R}^3$), by
\begin{equation}
{\mathcal F}\;:=\; \overline{{\mathcal F}^I\otimes{\mathcal
    F}^{I\!I}}, 
\label{eq:2.33}
\end{equation}
where ${\mathcal F}^I$ and ${\mathcal F}^{I\!I}$ are the field
algebras of the two reservoirs (see (\ref{eq:2.14})), and the closure is taken in the
operator norm.

A {\it contact} or {\it tunnelling junction} between the two
reservoirs is described 
in terms of a {\it perturbed Hamiltonian}, $H$, of the coupled
system. The operator $H$ has the form
\begin{equation}
H\;=\;H^0+W(\Lambda^I,\Lambda^{I\!I})
\label{eq:2.34}
\end{equation}
where $W(\Lambda^I,\Lambda^{I\!I})$ is a bounded, selfadjoint operator
on ${\mathcal H}$ for each choice of $\Lambda^I$ and
$\Lambda^{I\!I}$. We shall {\it always} require the following
assumption.

\vspace{.5cm}

{\bf (A3)~ Existence of the thermodynamic limit of the contact
  interaction}.
{\it 
\begin{equation}
\displaystyle\mathop{n-\lim}_{\substack{\Lambda^I\nearrow\infty \\
\Lambda^{I\!I}\nearrow\infty}}
~W\left(\Lambda^I, \Lambda^{I\!I}\right) \;=:\;W
\label{eq:2.35}
\end{equation}
exists (as a selfadjoint operator in ${\mathcal F}$).
}\\

\indent
Let $\alpha_t^{\Lambda^I}$ and $\alpha_t^{\Lambda^{I\!I}}$ be the time
evolutions of the reservoirs before they are brought into contact; see
eq.~(\ref{eq:2.1}). The
time evolution of operators in $B({\mathcal H}^I) \otimes B({\mathcal
  H}^{I\!I})$ is then given by $\alpha_t^{\Lambda^I} \otimes
\alpha_t^{\Lambda^{I\!I}}$ and is generated by the Hamiltonian $H^0$
introduced in (\ref{eq:2.32}). After the interaction
$W(\Lambda^I,\Lambda^{I\!I})$ has been turned on the time evolution of
operators in the Heisenberg picture is given by
\begin{equation}
\alpha_t^{I\cup I\!I} (a) \;=\;e^{i(tH/\hbar)}\,a\;e^{-i(tH/\hbar)}, 
\label{eq:2.36}
\end{equation}
with $H$ as in (\ref{eq:2.34}), for $a\in B({\mathcal H}^I) \otimes
B({\mathcal H}^{I\!I})$. 

It follows from assumptions (A1) and (A3), Eqs. (\ref{eq:2.16}) and
(\ref{eq:2.35}), that the thermodynamic limit of the time evolution of
the coupled reservoirs exists: For arbitrary $a\in{\mathcal F}$ the limits
\begin{equation}
\alpha_t^0(a)\;=\; 
\displaystyle\mathop{n-\lim}_{\substack{\Lambda^I\nearrow\infty \\
\Lambda^{I\!I}\nearrow\infty}} \;\alpha_t^{\Lambda^I} \otimes
\alpha_t^{\Lambda^{I\!I}} (a)
\label{eq:2.37}
\end{equation}
and
\begin{equation}
\alpha_t(a) \;=\;\displaystyle\mathop{n-\lim}_{\substack{\Lambda^I\nearrow\infty
\\ \Lambda^{I\!I}\nearrow\infty}}\;\alpha_t^{I\cup I\!I}\;(a)
\label{eq:2.38}
\end{equation}
exist, and the convergence is {\it uniform} on arbitrary compact
intervals of the time axis.

Given (\ref{eq:2.37}), (\ref{eq:2.38}) follows by using the {\it
  Lie-Schwinger series} for $\alpha_t (\alpha_{-t}^0 (a))$.\\
\noindent
We distinguish between {\it different types of contacts or junctions} between 
the two reservoirs, according to symmetry properties of the contact interactions $W(\Lambda^I,
\Lambda^{I\!I})$.

\vspace{.5cm}

{\bf (J1)~ Thermal contacts}. ~The interaction
$W(\Lambda^I,\Lambda^{I\!I})$ commutes with {\it all} the conservation
laws $\{ Q_j^{\Lambda^I} \otimes \id \}_{j=1}^{M^I}$ and $\{\id
\otimes  Q_j^{\Lambda^{I\!I}}\}_{j=1}^{M^{I\!I}}$ of the reservoirs,
i.e.,
\begin{equation}
\varphi_\us^{\Lambda^I}\bigl( W\bigl(
\Lambda^I,\Lambda^{I\!I}\bigr)\bigr) \;=\;  
\varphi_\us^{\Lambda^{I\!I}}
\bigl( W\bigl(\Lambda^I,\Lambda^{I\!I}\bigr)\bigr) \;=\;
W\bigl(\Lambda^I, \Lambda^{I\!I}\bigr),
\label{eq:2.40}
\end{equation}
for {\it arbitrary} $\Lambda^I$ and $\Lambda^{I\!I}$, where
  $\varphi_\us^{\Lambda^I}$ is a shorthand notation for
  $\varphi_\us^{\Lambda^I} \otimes {\rm id}$, and
  $\varphi_\us^{\Lambda^{I\!I}}$ stands for ~${\rm id} \otimes
   \varphi_\us^{\Lambda^{I\!I}}$; (see eq.~(\ref{eq:2.3})). It follows
   from (\ref{eq:2.34}) and (\ref{eq:2.2}) that the operators $\{
   Q_j^{\Lambda^I} \otimes \id\}$ and $\{\id\otimes
   Q_i^{\Lambda^{I\!I}}\}$ are conservation laws of the perturbed
   dynamics. 

By assumption (A1), the limits 
\begin{equation}
\displaystyle\mathop{n-\lim}_{\Lambda^r\nearrow\infty}\;
\varphi_\us^{\Lambda^r} (a) \;=:\; \varphi_\us^r (a)
\label{eq:2.41}
\end{equation}
exist for $r=I$ or $I\!I$ and for all $a\in{\mathcal F}$. By assumption
(A3) and (\ref{eq:2.40}) it follows that
\begin{equation}
\varphi_\us^I (W)\;=\;\varphi_\us^{I\!I}(W)\;=\;W,
\label{eq:2.42}
\end{equation}
in the thermodynamic limit.

Energy appears to be the only thermodynamic quantity that can be exchanged through a
thermal contact.

\vspace{.5cm}
{\bf (J2)~ Tunnelling junctions}.~ There are $m \leq {\rm
  min}\,(M^I,M^{I\!I})$ linear combinations,
$\widetilde{Q}_1^{\Lambda^I},\ldots,  \widetilde{Q}_m^{\Lambda^I}$ and
$\widetilde{Q}_1^{\Lambda^{I\!I}}, \ldots,
\widetilde{Q}_m^{\Lambda^{I\!I}}$, of conservation laws of the two
reservoirs with the property that the operators
\begin{equation}
Q_j^{I\cup I\!I}\;:=\;\widetilde{Q}_j^{\Lambda^I} \otimes \id + \id
\otimes\widetilde{Q}_j^{\Lambda^{I\!I}}, 
\label{eq:2.43}
\end{equation}
$j=1,\ldots,m$, are {\it conservation laws} of the {\it perturbed
  dynamics} generated by the Hamiltonian $H$ of
eq.~(\ref{eq:2.34}). Without loss of generality, we may assume that
$\widetilde{Q}_j^{\Lambda^I}= Q_j^{\Lambda^I}$ and
$\widetilde{Q}_j^{\Lambda^{I\!I}}=Q_j^{\Lambda^{I\!I}}$, for
$j=1,\ldots,m$. Of course, there may be further conservation laws of
the reservoirs,
$Q_i^{\Lambda^I}\otimes \id$, $\id \otimes Q_j^{\Lambda^{I\!I}}$,
for some $i>m$ and/or some $j>m$, which are conservation laws of the
perturbed dynamics. ``Leaky junctions'' are contacts where the
interaction $W(\Lambda^I,\Lambda^{I\!I})$ {\it violates some} of the
conservation laws $Q_i^{\Lambda^I} \otimes\id$ and/or $\id\otimes
Q_j^{\Lambda^{I\!I}}$, $i,j>m$. For convenience, we shall sometimes
assume that the operators $Q_j^{I\cup I\!I}$, $j=1,\ldots,m$, are the
{\it only} conservation laws of the perturbed dynamics, and
$M^I=M^{I\!I}=m$. Let $\us = (s_1,\ldots,s_m,0,\ldots,0)$. We define
\begin{equation}
\varphi_\us^{I\cup I\!I} (a) \;:=\; \varphi_\us^{\Lambda^I} \otimes
\varphi_\us^{\Lambda^{I\!I}} (a),
\label{eq:2.44}
\end{equation}
for $a\in B({\mathcal H}^I) \otimes B({\mathcal H}^{I\!I})$, and
\begin{equation}
\varphi_\us (a) \;:=\;
\displaystyle\mathop{n-\lim}_{\substack{\Lambda^I\nearrow\infty \\
\Lambda^{I\!I}\nearrow\infty}} \;\varphi_\us^{I\cup I\!I} (a),
\label{eq:2.45}
\end{equation}
for $a\in{\mathcal F}$; see assumptions (A1), eq.~(\ref{eq:2.17}). 

Tunnelling junctions can then be characterized by the requirement that
\begin{equation}
\varphi_\us^{I\cup I\!I} \bigl( W\bigl(
\Lambda^I,\Lambda^{I\!I}\bigr)\bigr) \;=\; W \bigl(
\Lambda^I,\Lambda^{I\!I}\bigr), 
\label{eq:2.46}
\end{equation}
for arbitrary $\Lambda^I,\Lambda^{I\!I}$, and hence, using
(\ref{eq:2.17}) and (\ref{eq:2.35}), we find that, in the
thermodynamic limit, 
\begin{equation}
\varphi_\us (W)\;=\;W.
\label{eq:2.47}
\end{equation}
As an {\it initial state} of a tunnelling junction we shall usually
choose a state $\omega$ close to a tensor product state,
$\omega_{\beta^I,\umu^I}^{\Lambda^I} \otimes
\omega_{\beta^{I\!I},\umu^{I\!I}}^{\Lambda^{I\!I}}$, of two equilibrium
states of the uncoupled reservoirs, where $\beta^I,\umu^I$ and
$\beta^{I\!I},\umu^{I\!I}$ are arbitrary, (with $\umu^I\in {\mathcal
  M}^I, \umu^{I\!I}\in{\mathcal M}^{I\!I}$).

Two reservoirs joined by a tunnelling junction can exchange energy and\break
``charge'' (as measured by the conservation laws $Q_j^{\Lambda^I}
\otimes \id$, $\id \otimes Q_j^{\Lambda^{I\!I}}$, $j=1,\ldots,m$), or
leak some ``charge'' corresponding to $Q_j^{\Lambda^I}\otimes\id$, or
to $\id\otimes Q_j^{\Lambda^{I\!I}}$, for some $j>m$.

\medskip

{\bf Energy current}.~ The operator corresponding to a measurement of
the {\it gain of internal energy per second} of reservoir $r$, with
$r=I$ or $I\!I$, at time $t$ is conveniently defined in the
Heisenberg picture by
\begin{eqnarray}
P^r(t) &:=& \frac{d}{dt}~\alpha_t^{I\cup I\!I} (H^r)\nonumber \\ 
&=& \frac{i}{\hbar}~\alpha_t^{I\cup I\!I} ([H,H^r]),
\label{eq:2.48}
\end{eqnarray}
where $\alpha_t^{I\cup I\!I}$ is as in (\ref{eq:2.36}), and
$H^r=H^I\otimes\id$ or $=\id\otimes H^{I\!I}$, for $r=I$ or $I\!I$,
respectively. By (\ref{eq:2.32}) and (\ref{eq:2.34}), 
\begin{equation}
P^r(t)\;=\;\frac i\hbar~ \alpha_t^{I\cup I\!I} \bigl(\bigl[
W(\Lambda^I,\Lambda^{I\!I}),H^r\bigr]\bigr). 
\label{eq:2.49}
\end{equation}
By (\ref{eq:2.16}), (\ref{eq:2.35}) and (\ref{eq:2.38}),
the operator corresponding to the energy gain per
second of reservoir $r$ has a thermodynamic limit given by 
\begin{equation}
P^r(t)\;=\;-\,\frac{d}{ds}~ \alpha_t\,\bigl(
\alpha_s^r(W)\bigr)\big|_{s=0}, 
\label{eq:2.50}
\end{equation}
where $\alpha_s^r$ is the time evolution of reservoir $r$ in the
thermodynamic limit, in the absence of any contacts. It follows from
(\ref{eq:2.49}) and (\ref{eq:2.50}) that 
\begin{eqnarray}
P^I(t) + P^{I\!I}(t) &=& \frac i\hbar~ \alpha_t^{I\cup
  I\!I}\,\bigl(\bigl[ W(\Lambda^I, \Lambda^{I\!I}),
H^0\bigr]\bigr)\nonumber \\
&=& \frac i\hbar~ \alpha_t^{I\cup I\!I} \bigl(\bigl[ W(
\Lambda^I,\Lambda^{I\!I}),H\bigr]\bigr)\nonumber \\
&=&-\,\frac{d}{dt}~\alpha_t^{I\cup I\!I} \bigl(
W(\Lambda^I,\Lambda^{I\!I})\bigr), 
\label{eq:2.51}
\end{eqnarray}
where $H^0$ and $H$ are as in eqs.~(\ref{eq:2.32}), (\ref{eq:2.34}),
and, in the thermodynamic limit,
\begin{equation}
P^I(t) + P^{I\!I}(t) \;=\;-\,\frac{d}{dt}~ \alpha_t (W),
\label{eq:2.52}
\end{equation}
with $W\in{\mathcal F}$.

We observe that if $\omega$ is an {\it arbitrary time-translation
  invariant state} of the coupled system, we have that
\begin{equation}
\omega\,\bigl( P^I(t) + P^{I\!I}(t)\bigr)\;=\;0,
\label{eq:2.53}
\end{equation}
for all times.

\medskip

{\bf Charge current}.~ The operator corresponding to a measurement of
the gain of charge $Q_j^{r}$ per second at time $t$, in
reservoir $r$, is conveniently defined by
\begin{eqnarray}
I_j^r(t) &=& \frac{d}{dt}~ \alpha_t^{I\cup I\!I} \,(Q_j^r) \nonumber\\
&=& \frac i\hbar~ \alpha_t^{I\cup I\!I}\,([H,Q_j^r]),
\label{eq:2.54}
\end{eqnarray}
for $j=1,\ldots,m$, $r=I,I\!I$, with $Q_j^I := Q_j^{\Lambda^I}\otimes
\id$, and $Q_j^{I\!I} := \id\otimes Q_j^{\Lambda^{I\!I}}$. Since $H$
is given by
\[
H\;=\;H^I\otimes\id\;+\;\id\otimes H^{I\!I}\;+\;W
\bigl( \Lambda^I,\Lambda^{I\!I}\bigr),
\]
and since $H^I\otimes\id$ and $\id\otimes H^{I\!I}$ commute with
$Q_i^r$, it follows that
\begin{equation}
I_j^r(t)\;=\;\frac i\hbar~ \alpha_t^{I\cup I\!I}\, \bigl(\bigl[
W\,(\Lambda^I,\Lambda^{I\!I}), Q_j^r\bigr]\bigr).
\label{eq:2.55}
\end{equation}
In the thermodynamic limit,
\begin{equation}
I_j^r(t)\;=\;-\,\frac 1 \hbar~\frac{\partial}{\partial\, s_j}~
\alpha_t \bigl( \varphi_\us^r (W)\bigr)\big|_{\us=0}
\label{eq:2.56}
\end{equation}
with $\varphi_\us^r$ as in (\ref{eq:2.41}); $r=I,I\!I$,
$j=1,\ldots,M^r$. 
Recall that, for $j=1,\ldots,m$, the operators $Q_j^{I\cup I\!I}=Q_j^{\Lambda^I}\otimes\id
+\id\otimes Q_j^{\Lambda^{I\!I}}$ are conservation laws of the
perturbed dynamics, see (\ref{eq:2.46}), and therefore
\begin{eqnarray}
I_j^I(t)\,+\,I_j^{I\!I}(t) &=& \frac i \hbar~ \alpha_t^{I\cup I\!I}
\bigl(\bigl[ W\left(\Lambda^I,\Lambda^{I\!I}\right), Q_j^{I\cup
  I\!I}\bigr]\bigr)=0,
\label{eq:2.57}
\end{eqnarray}
for $j=1,\ldots, m$; 
this can be transferred to the thermodynamic limit. Apparently,
charge lost by one reservoir is gained by the other one.

For $j>m$, (\ref{eq:2.57}) does {\it not} hold in general and the
operators $I_j^r (t), r=I,I\!I$, describe the leakage of charge
$Q_j^r$ at the junction.

\subsection{Connections with thermodynamics} \label{subsec-2.3}

We start by recalling the {\bf 1$^{{\bf s}{\bf t}}$ and
  2$^{{\bf n}{\bf d}}$
  law of thermodynamics}.

For the reservoir $r$, the first and second law of thermodynamics can be
summarized in the equation
\begin{equation}
d
U^{\Lambda^r}\;=\;T^r\,dS^{\Lambda^r}\;+\;\umu^r\,\cdot\,d\uq^{\Lambda^r}
- p^r\,dV^r,
\label{eq:2.58}
\end{equation}
where $U^{\Lambda^r}$ is the expectation value of the Hamiltonian
$H^r$ in a state of reservoir $r$ {\it close to}, or {\it in} thermal
equilibrium, i.e., $U^{\Lambda^r}$ is the {\it internal energy} of the
reservoir $r$; $T^r$ the {\it temperature}; $S^{\Lambda^r}$ the {\it
  entropy}; $q_j^{\Lambda^r}$ is the expectation value of the charge
$Q_j^r$, $j=1,\ldots,M^r$, in the state describing the reservoir; $p^r$ is the
{\it pressure}, and $V^r={\rm vol}(\Lambda^r)$ the volume. The
differential ``$d$'' indicates that we consider the variation of
$U^{\Lambda^r}$, $S^{\Lambda^r}$, etc. under small, reversible changes of the
state of reservoir $r$ (which may include small changes of the region
$\Lambda^r$). 

We shall be interested in studying small, slow changes in time of the
state of reservoirs $I$ and $I\!I$, at approximately fixed values of 
the thermodynamic parameters $T^r, \umu^r$ and $p^r$, brought about by
opening a {\it contact} or {\it junction} between the two
reservoirs. Then $U^{\Lambda^r}, S^{\Lambda^r},\ldots$~ are {\it
  time-dependent}, and (\ref{eq:2.58}) becomes
\begin{equation}
\dot{U}^{\Lambda^r}\;=\;T^r\dot{S}^{\Lambda^r}\,+\,\umu^r \cdot
\dot{\uq}^{\Lambda^r} - p^r\,\dot{V}^r, 
\label{eq:2.59}
\end{equation}
where the ``dots'' indicate time derivatives. 
By (\ref{eq:2.48}), the energy gain per second, $\dot{U}^{\Lambda^r}$,
of the reservoir $r$ is given by
\begin{equation}
\dot{U}^{\Lambda^r} (t)\;=\;\omega^{I\cup I\!I} \left( P^r(t)\right),
\label{eq:2.60}
\end{equation}
and the gain in the j$^{\rm th}$ charge per second by
\begin{equation}
\dot{q}_j^{\Lambda^r} (t) \;=\;\omega^{I\cup I\!I} \left( I_j^r
  (t)\right), 
\label{eq:2.61}
\end{equation}
see (\ref{eq:2.54}), where $\omega^{I\cup I\!I}$ is the state of the
system consisting of the two reservoirs. By (\ref{eq:2.59}), the
change in entropy per second of reservoir $r$ is given by $(\beta^r :=
1/T^r)$
\begin{equation}
\dot{S}^{\Lambda^r}\;=\;\beta^r\left( \dot{U}^{\Lambda^r} - \umu^r
  \cdot \dot{\uq}^{\Lambda^r} +\, p^r \dot{V}^r\right). 
\label{eq:2.62}
\end{equation}
We define the {\it entropy production rate}, ${\mathcal E}^{I\cup
  I\!I}$, by 
\begin{equation}
{\mathcal E}^{I\cup I\!I}\;:=\;\dot{S}^{\Lambda^I} +\,
\dot{S}^{\Lambda^{I\!I}} .
\label{eq:2.63}
\end{equation}
The main property of ${\mathcal E}^{I\cup I\!I}$ is {\it its sign}:
thermodynamic systems should exhibit positive entropy production,
\begin{equation}
{\mathcal E}^{I\cup I\!I}\;\geq\;0,
\label{eq:2.64}
\end{equation}
in the limit where $\Lambda^I\nearrow\infty$ and $\Lambda^{I\!I}\nearrow\infty$. 
In [Ru2], Ruelle has proven (\ref{eq:2.64}) for the special case of
thermal contacts between infinitely large reservoirs. Below, we shall
derive (\ref{eq:2.64}), under more general conditions, from the
positivity of ``relative entropy''; see also [JP1].

We consider a situation in which the state of the system consisting of
reservoirs $I$ and $I\!I$, {\it before} a contact or junction is
opened, is given by the tensor product of two equilibrium states
\begin{equation}
\omega^{I\cup I\!I} (a)\;=\;\omega_{\beta^I,\umu^I}^{\Lambda^I}
\otimes \omega_{\beta^{I\!I},\umu^{I\!I}}^{\Lambda^{I\!I}} (a),
\label{eq:2.66}
\end{equation}
for any operator $a \in B ({\mathcal H}^{\Lambda^I}) \otimes B
({\mathcal H}^{\Lambda^{I\!I}})$, where the equilibrium states
$\omega_{\beta^r,\umu^r}^{\Lambda^r}$ have 
been defined in (\ref{eq:2.10}). The state $\omega^{I\cup I\!I}$
is invariant under the unperturbed time evolutions
$\alpha_t^{\Lambda^r}$, $r=I,I\!I$, of the reservoirs.

At some time $t_0$, the contact between the reservoirs is opened, and
we are interested in the evolution of the state $\omega^{I\cup I\!I}$
under the {\it perturbed} time evolution, $\alpha_t^{I\cup I\!I}$,
introduced in (\ref{eq:2.34}), (\ref{eq:2.36}). In particular, we are
interested in calculating the rate of energy gain, or loss,
$\dot{U}^{\Lambda^r} (t)$, the gain or loss of charge $j$,
$\dot{q}_j^{\Lambda^r} (t), j=1,\ldots,M^r$, per second and the
entropy production rate ${\mathcal E}^{I\cup I\!I}(t)$, under the
perturbed time evolution, in the state $\omega^{I\cup I\!I}$. 
By eqs.~(\ref{eq:2.48}) and (\ref{eq:2.49}), 
\begin{eqnarray}
\dot{U}^{\Lambda^r} (t) &=& \omega^{I\cup I\!I} (P^r(t)) \nonumber\\
&=& \frac i\hbar~ \omega^{I\cup I\!I} \bigl( \alpha_t^{I\cup I\!I}
\bigl(\bigl[ W(\Lambda^I,\Lambda^{I\!I}), H^r\bigr]\bigr)\bigr)
\label{eq:2.67}
\end{eqnarray}
and, by (\ref{eq:2.54}),
\begin{eqnarray}
\dot{q}_j^{\Lambda^r} (t) &=& \omega^{I\cup I\!I} \bigl( I_j^r
(t)\bigr) \nonumber\\
&=& \frac i\hbar~ \omega^{I\cup I\!I} \bigl( \alpha_t^{I\cup I\!I}
\bigl(\bigl[ W(\Lambda^I,\Lambda^{I\!I}),Q_j^r\bigr]\bigr)\bigr).
\label{eq:2.68}
\end{eqnarray}
By assumption (A2), see (\ref{eq:2.19}), the states $\omega^{I\cup
  I\!I}$ have a thermodynamic limit
\begin{equation}
\omega^0(a)\;=\;\lim_{\substack{\Lambda^I\nearrow\infty \\
  \Lambda^{I\!I}\nearrow\infty}}\; \omega^{I\cup I\!I} (a),
\label{eq:2.69}
\end{equation}
for $a\in {\mathcal F}={\mathcal F}^I \otimes {\mathcal
  F}^{I\!I}$. It follows from this property, from assumption (A3), and from equations (\ref{eq:2.38}),
(\ref{eq:2.50}), and (\ref{eq:2.56}), that the quantities
$\dot{U}^{\Lambda^r}(t)$ and $\dot{q}_j^{\Lambda^r}(t)$ have
thermodynamic limits
\begin{eqnarray}
{\mathcal P}^r(t) &:=& \lim_{\substack{\Lambda^I\nearrow\infty \\
  \Lambda^{I\!I}\nearrow\infty}}\; \dot{U}^{\Lambda^r}(t) \nonumber\\
&=&- \frac{d}{ds}~ \omega^0\bigl(\alpha_t\bigl(\alpha_s^r
(W)\bigr)\bigr)\big|_{s=0} 
\label{eq:2.70}
\end{eqnarray}
and
\begin{eqnarray}
{\mathcal J}_j^r(t) &:=& \lim_{\substack{\Lambda^I\nearrow\infty \\
  \Lambda^{I\!I}\nearrow\infty}}\; \dot{q}_j^{\Lambda^r} (t)\nonumber\\
&=& - \frac 1\hbar~\frac{\partial}{\partial s_j}~ \omega^0
\bigl(\alpha_t\bigl(\varphi_\us^r(W) \bigr)\bigr)\big|_{\us=0}~.
\label{eq:2.71}
\end{eqnarray}
These limits are {\it uniform} on compact intervals of the time
axis. By (\ref{eq:2.51}) and (\ref{eq:2.57}), 
\begin{equation}
{\mathcal P}^I (t) + {\mathcal P}^{I\!I}(t)\;=\;- \frac{d}{dt}~
\omega^0 \bigl( \alpha_t (W)\bigr)
\label{eq:2.72}
\end{equation}
and
\begin{equation}
{\mathcal J}_j^I(t) + {\mathcal J}_j^{I\!I}(t)\;=\;0,
\label{eq:2.73}
\end{equation}
for $j=1,\ldots,m$.

Next, we study the entropy production rate for finite reservoirs. Let 
\begin{equation}
\rho^r\;:=\;\left( \Xi_{\beta^r,\umu^r}^{\Lambda^r}\right)^{-1}\;{\rm
  exp} -\beta^r \left[ H^r-\umu^r\cdot\uQ^{\Lambda^r}\right]
\label{eq:2.74}
\end{equation}
be the density matrix corresponding to the equilibrium state,
$\omega_{\beta^r, \umu^r}^{\Lambda^r}$, for the reservoir~$r$; see
(\ref{eq:2.10}). Then 
\begin{equation}
-\ln\, \rho^r\;=\;\beta^r\left[ H^r-\umu^r\cdot\uQ^{\Lambda^r}\right] -
\beta^r G\left( \beta^r, \umu^r, V^r\right)\;\cdot\;\id~,
\label{eq:2.75}
\end{equation}
where the thermodynamic potential $G$ is as in (\ref{eq:2.9}). If
the confinement region $\Lambda^r$ is kept constant in time, so that
$\dot{V}^r=0$, then it follows from (\ref{eq:2.59}), (\ref{eq:2.60}),
(\ref{eq:2.61}) and (\ref{eq:2.75}) that
\begin{eqnarray}
\dot{S}^{\Lambda^r} &=&\beta^r \left(\dot{U}^{\Lambda^r}-
  \umu^r\cdot\dot{\uq}^{\Lambda^r}\right)\nonumber\\
&=& -\frac{d}{dt}~ \omega^{I\cup I\!I} \left(\alpha_t^{I\cup
    I\!I}\left( \ln \rho^r\right)\right)\nonumber\\
&=&-\frac{d}{dt}~{\rm tr}\,\left( \rho^I\otimes \rho^{I\!I}\,
  \alpha_t^{I\cup I\!I} \left( \ln \rho^r\right)\right)~.
\label{eq:2.76}
\end{eqnarray}
Thus, by (\ref{eq:2.63}), 
\begin{eqnarray}
\dot{S}^{I\cup I\!I}(t) &:=& {\mathcal E}^{I\cup I\!I}(t)\nonumber\\
&=&-\frac{d}{dt}~ {\rm tr}\,\left( \rho^I\otimes
  \rho^{I\!I}\,\alpha_t^{I\cup I\!I}\,\left( \ln \rho^I\otimes
    \rho^{I\!I}\right)\right)~. 
\label{eq:2.77}
\end{eqnarray}
By (\ref{eq:2.70}) and (\ref{eq:2.71}) this quantity has a thermodynamic
limit
\begin{equation}
{\mathcal E}(t)\;=\sum_{r=I,I\!I}\, \beta^r \left[ {\mathcal
    P}^r(t)-\umu^r \cdot {\flow}^r(t)\right]~.
\label{eq:2.78}
\end{equation}
Integrating (\ref{eq:2.77}) in time, we find that
\begin{equation}
S^{I\cup I\!I}(t)-S^{I\cup I\!I}(0)
=\;-{\rm tr}\,\left( \rho^I\otimes \rho^{I\!I} \left[ \alpha_t^{I\cup
      I\!I} \left(\ln \rho^I \otimes \rho^{I\!I}\right)-\ln \rho^I\otimes
    \rho^{I\!I}\right]\right)~.
\label{eq:2.79}
\end{equation}
This equation shows that $S^{I\cup I\!I}(t)-S^{I\cup I\!I}(0)$ is
nothing but the {\it relative entropy} of the density matrix
$\alpha_t^{I\cup I\!I}(\rho^I\otimes \rho^{I\!I})$ with respect to the
density matrix $\rho^I\otimes \rho^{I\!I}$; see e.g. [BR, vol~II] for a
definition of 
relative entropy, which differs from ours by the sign, and [JP1] for
similar, independent considerations. If $A$ is a non-negative matrix
and $B$ is a strictly positive matrix then 
\begin{equation}
- {\rm tr}\, \left( A \ln B - A\ln A\right) \;\geq\; {\rm tr}\, (A-B),
\label{eq:2.80}
\end{equation}
see Lemma 6.2.21 of [BR, vol~II]. Setting $A=\rho^I\otimes \rho^{I\!I}$ and
$B=\alpha_t^{I\cup I\!I} (\rho^I\otimes \rho^{I\!I})$, we find that
\begin{equation}
S^{I\cup I\!I}(t)-S^{I\cup I\!I}(0) \;\geq\;{\rm tr}\left(\rho^I\otimes \rho^{I\!I}
  -\alpha_t^{I\cup I\!I}(\rho^I\otimes \rho^{I\!I})
 \right)\;=\;0, 
\label{eq:2.81}
\end{equation}
by the unitarity of time evolution and the cyclicity of the trace. It
follows that 
\begin{equation}
\frac 1T~\int_0^T {\mathcal E}^{I\cup I\!I}(t)\,{\rm dt}\;=\; \frac
1T~\left( S^{I\cup I\!I}(T)-S^{I\cup I\!I}(0)\right)\;\geq\;0,
\label{eq:2.82}
\end{equation}
and this inequality remains obviously valid in the thermodynamic
limit:
\begin{equation}
\frac 1T~\int_0^T {\mathcal E}(t)\;{\rm dt}\;\geq\;0~.
\label{eq:2.83}
\end{equation}
Thus, if the limit
\begin{equation}
\lim_{t\to\infty}~{\mathcal E}(t)\;=:\;{\mathcal E}
\label{eq:2.84}
\end{equation}
exists then
\begin{equation}
{\mathcal E}\;=\;\lim_{T\to\infty}~\frac 1T~ \int_0^T {\mathcal
  E}(t)\,{\rm dt}\;\geq\; 0~,
\label{eq:2.85}
\end{equation}
i.e., the {\it entropy production rate} ${\mathcal E}$, in the
thermodynamic limit, is non-negative, as time $t$ tends to $\infty$;
see [Ru1]. In Sect.~5, we shall study examples where ${\mathcal E}$ is
{\it strictly positive}.

Let us assume that the entropy production rate
${\mathcal E}(t)$ converges as $t\to\infty$. It follows from \eqref{eq:2.83} and \eqref{eq:2.85} that
it is {\it nonnegative}.

Let us set
\begin{equation}
{\mathcal P}\;:=\;{\mathcal P}^I\;=\;-{\mathcal P}^{I\!I}
\label{eq:2.98}
\end{equation}
and, for $j=1,\ldots,m$,
\begin{equation}
{\mathcal J}_j\;:=\;{\mathcal J}_j^I\;=\;-{\mathcal J}_j^{I\!I}.
\label{eq:2.99}
\end{equation}
In the case where each reservoir has precisely $m$
conservation laws $\{ Q_j^r\}_{j=1}^m$, $r=I,I\!I$ (that is,
$M^I=M^{I\!I}=m$ and
$\{Q_j^I\otimes\id+\id\otimes Q_j^{I\!I}\}_{j=1}^m$ are conservation
laws of the coupled system), nonnegativity of the entropy production rate
implies that 
\begin{equation}
{\mathcal E}\;=\;\left(\beta^I-\beta^{I\!I}\right)\,
{\mathcal P}-\left(\beta^I\umu^I-\beta^{I\!I}\umu^{I\!I}\right)\,\cdot
\flow\;\geq\;0~.
\label{eq:2.100}
\end{equation}
The currents ${\mathcal J}_j^r$
vanish for thermal contacts, and \eqref{eq:2.100} shows that energy is
transferred from the hotter to the colder
reservoir --- as expected.

The thermoelectric situation corresponds to $M^I=M^{I\!I}=m=1$, and $Q_1^r=N^r$ is the
particle number operator. For identical temperatures but different chemical potentials,
\eqref{eq:2.100} shows that particles are transferred from the reservoir with the higher chemical
potential to the reservoir with the lower chemical potential. Notice that energy may flow from the
colder reservoir to the hotter one when the chemical potentials are different (consider e.g.\
$\beta^I\mu^I\gg \beta^{I\!I}\mu^{I\!I}$ but $\beta^I>\beta^{I\!I}$).

Also interesting is the case of {\it adiabatic} thermal contacts between two
reservoirs, i.e., without heat exchange. A general discussion of systems with
time-dependent interactions confined to time-dependent regions is given in
[FMSU], and will be elaborated upon in a forthcoming paper.

\subsection{Existence of stationary states in the thermodynamic limit}
\label{subsec-2.4} 

The above considerations, and in particular (\ref{eq:2.70}),
(\ref{eq:2.71}), and (\ref{eq:2.84}), suggest to study the question
whether the infinite-volume states
\begin{equation}
\omega_t(a)\;:=\; \omega^0\left( \alpha_t(a)\right),~~ a\in{\mathcal
  F}~,
\label{eq:2.86}
\end{equation}
have a limit, as $t\to\infty$.

The state $\omega^0$, defined in (\ref{eq:2.69}), is obviously invariant under the
unperturbed 
time evolution $\alpha_t^0$ defined in (\ref{eq:2.37}). Thus
\begin{equation}
\omega_t(a)\;=\;\omega^0\left(\alpha_{-t}^0\left(\alpha_t(a)
  \right)\right),~ a\in {\mathcal F}.
\label{eq:2.87}
\end{equation}
A sufficient condition for the existence of a {\it stationary} (i.e.,
time-translation invariant) {\it limiting state},
\begin{equation}
\omega_{\rm stat}(a)\;=\;\lim_{t\to\infty}\; \omega_t(a),~~
a\in{\mathcal F}~,
\label{eq:2.88}
\end{equation}
is given in\\

{\bf (A4)~ Existence of a scattering endomorphism}.

{\it The limits
\begin{equation}
\sigma_\pm (a)\;=\;\displaystyle\mathop{n-\lim}_{t\to\pm\infty}\;
\alpha_{-t}^0\left(\alpha_t(a)\right) 
\label{eq:2.89}
\end{equation}
exist, for all $a\in{\mathcal F}$, and define $^*${\it endomorphisms}
of ${\mathcal F}$, i.e., $\sigma_\pm$ are homomorphisms of the
$C^*$-algebra ${\mathcal F}$ with the property that
$\sigma_\pm(a)^*=\sigma_\pm(a^*)$, for all $a\in{\mathcal F}$.
}\\

\noindent
The usefulness of these so-called {\it scattering (or M{\o}ller)
endomorphisms} has first been recognized in [He, Rob]; interesting
examples have been constructed in [BM]. In the context of thermal
contacts and tunnelling junctions, they have first been used in [DFG];
see also [Ma].

It is important to note that scattering endomorphisms do not exist in {\it finite} volume, because the free and the
perturbed time evolutions of the two reservoirs are generated by Hamiltonians
\begin{eqnarray*}
H^0 &=& H^I\otimes\id +\id \otimes H^{I\!I},\\
H &=& H^0+W(\Lambda^I, \Lambda^{I\!I}),
\end{eqnarray*}
see (\ref{eq:2.32}) and (\ref{eq:2.34}), with {\it
  pure-point} spectra when $\Lambda^I$ and $\Lambda^{I\!I}$ are
compact. It is thus natural to wonder about the meaning of
scattering endomorphisms for large but finite reservoirs.
Let us sketch the answer to this question. We fix an
arbitrarily small, but positive number $\varepsilon$. For every operator
$a\in{\mathcal F}$, there exist compact regions
$\Lambda^r(\varepsilon, a)$, $r=I,I\!I$, and an operator
$a_\varepsilon \in B ({\mathcal H}^{\Lambda^I}) \otimes B({\mathcal
  H}^{\Lambda^{I\!I}})$, with $\Lambda^r=\Lambda^r(\varepsilon, a)$,
$r=I,I\!I$, such that
\[
\Vert a-a_\varepsilon\Vert\;<\;\frac \varepsilon 4 ~.
\]
Then, by (\ref{eq:2.88}) and (\ref{eq:2.89}), there is some
$T(\varepsilon, a) < \infty$ such that
\[
\big| \omega_{\rm stat}(a) - \omega_t(a_\varepsilon)\big| \;<\;\frac
\varepsilon 2 ~,
\]
for all $t>T(\varepsilon,a)$. Assumption (A3), eq.~(\ref{eq:2.38}),
tells us that, for an arbitrary $T<\infty$, there are compact sets
$\Lambda^r(\varepsilon, a,T) \supseteq \Lambda^r(\varepsilon,a)$ such
that if $\Lambda^r\supset \Lambda^r(\varepsilon,a,T)$, $r=I,I\!I$,
then
\[
\Vert \alpha_t(a_\varepsilon) - \alpha_t^{I\cup I\!I}
(a_\varepsilon)\Vert \;<\;\frac \varepsilon 4 ~,
\]
for all $t\in [0,T]$. Finally, by assumption (A2), one can choose
$\Lambda^r(\varepsilon,a,T)$ so large that
\[
\big| \omega^{I\cup I\!I}\left( \alpha_t^{I\cup
    I\!I}(a_\varepsilon)\right) - \omega^0\left( \alpha_t^{I\cup I\!I}
  (a_\varepsilon)\right)\big| \;<\;\frac \varepsilon 4 ~,
\]
provided $\Lambda^r \supset \Lambda^r (\varepsilon, a, T)$,
$r=I,I\!I$, for {\it all} times $t \in [0,T]$. It follows that, for any $T$,
with 
$ 
0 \;<\;T (\varepsilon, a)\;<\;T\;<\;\infty~,
$ 
and for 
$
\Lambda^r\;\supset\;\Lambda^r(\varepsilon,a,T),~~ r\;=\;I,I\!I,
$  
\begin{equation}
\big| \omega_{\rm stat}(a)-\omega^{I\cup I\!I} \left(\alpha_t^{I\cup
    I\!I}(a_\varepsilon)\right)\big| \;<\;\varepsilon~,
\label{eq:2.90}
\end{equation}
for all times $t$, with $T(\varepsilon, a)<t<T$.

These simple considerations, combined with (\ref{eq:2.52}) and
(\ref{eq:2.56}), show that the energy-gain rates
$\dot{U}^{\Lambda^r}(t)$ and the currents $\dot{q}_j^{\Lambda^r}(t)$
of two very large, but finite reservoirs, $r=I,I\!I$, are 
well approximated by the energy-gain rates
\begin{equation}
{\mathcal P}^r \;:=\;\lim_{t\to\infty}\;{\mathcal P}^r(t)
\;=\;-\frac{d}{ds}~ \omega_{\rm stat} \left(\alpha_s^r
  (W)\right)\big|_{s=0} 
\label{eq:2.91}
\end{equation}
and the currents
\begin{equation}
{\mathcal J}_j^r\;:=\;\lim_{t\to\infty}\;{\mathcal J}_j^r(t) \;=\;
-\frac 1\hbar~ \frac{\partial}{\partial s_j}~ \omega_{\rm stat}
\left(\varphi_\us^r(W)\right) \big|_{\us=0}~ ,
\label{eq:2.92}
\end{equation}
respectively, for a large range of {\it sufficiently large}, but {\it
  not exceedingly large} times $t$; (see (\ref{eq:2.70}),
(\ref{eq:2.71})).

\vspace{.5cm} 
{\bf Remark}.~ It is usually much easier to prove that the limits
\begin{equation}
\sigma_\pm'(a)\;:=\;\displaystyle\mathop{n-\lim}_{t\to\pm\infty}\;
\alpha_{-t} \left(\alpha_t^0(a)\right)
\label{eq:2.93}
\end{equation}
exist and are operators in ${\mathcal F}$, for arbitrary $a\in{\mathcal
    F}$, rather than to establish the existence of the scattering
  endomorphisms $\sigma_\pm$ in (\ref{eq:2.89}). If the unperturbed
  dynamics of
the reservoirs is dispersive, (as for non-interacting,
non-relativistic electrons), one may hope to prove
(\ref{eq:2.93}) by using a simple Cook argument; see e.g. [He, Rob, CFKS]. If
both limits (\ref{eq:2.89}) and (\ref{eq:2.93}) exist then
\begin{equation}
\sigma_\pm\left(\sigma_\pm'(a)\right)\;=\;\sigma_\pm'\left(
  \sigma_\pm(a)\right)\;=\;a~ , 
\label{eq:2.94}
\end{equation}
i.e., $\sigma_\pm'$ is a left and right inverse of $\sigma_\pm$, and
hence $\sigma_\pm$ is a $^*${\it automorphism} of ${\mathcal F}$. This
will turn out to hold in the examples discussed in subsequent
sections.

\subsection{Uniqueness and stability properties of stationary
  states} \label{subsec-2.6} 

We first describe the property of return to equilibrium for a single
reservoir.  Let $\omega$ be a state on the field algebra of a single
reservoir, 
${\mathcal F}^r$, i.e., $\omega$ is a positive, linear functional on
${\mathcal F}^r$ normalized such that $\omega (\id)=1$. From ${\mathcal
  F}^r$ and $\omega$ one can construct a Hilbert space ${\mathcal
  H}_\omega$, a representation $\pi_\omega$ of ${\mathcal F}^r$ on
${\mathcal H}_\omega$, and a unit vector $\Omega\in{\mathcal
  H}_\omega$ (unique up to a phase) such that
\begin{equation}
{\mathcal H}_\omega\;=\; 
\overline{\left\{ \pi_\omega(a)\Omega\bigm| a\in{\mathcal F}^r\right\}},
\label{eq:2.23}
\end{equation}
where the closure is taken in the norm on ${\mathcal H}_\omega$, i.e.,
$\Omega$ is ``cyclic'' for $\pi_\omega ({\mathcal F}^r)$, 
and
\begin{equation}
\omega(a)\;=\;\big\langle \Omega, \pi_\omega (a)\,\Omega\big\rangle ,
\label{eq:2.24}
\end{equation}
where $\langle\cdot,\cdot\rangle$ is the scalar product on ${\mathcal
  H}_\omega$. This is the content of the {\it Gel'fand-Naimark-Segal}
(GNS) {\it construction}. If the state $\omega$ is time-translation
invariant then there exists a one-parameter unitary group $\left\{
  U_\omega (t) \bigm| t\in{\mathbb R}\right\}$ on ${\mathcal
  H}_\omega$ such that
\[
\pi_\omega ( \alpha_t (a)) \;=\; U_\omega (t)\,
\pi_\omega (a)\, U_\omega (t)^*,
\]
and
\begin{equation}
U_\omega (t)\, \Omega\;=\;\Omega~.
\label{eq:2.25}
\end{equation}
Under standard continuity assumptions on $U_\omega(t)$, we can summon
Stone's theorem to conclude that
\begin{equation}
U_\omega(t)\;=\; e^{it\,L_\omega/\hbar},
\label{eq:2.26}
\end{equation}
where the generator $L_\omega$ is a selfadjoint operator on ${\mathcal
  H}_\omega$ with $L_\omega\,\Omega=0$.


A state $\rho$ on ${\mathcal F}^r$ is called {\it normal} relative to
$\omega$ iff there exists a density matrix, $P$, on ${\mathcal
  H}_\omega$ such that
\begin{equation}
\rho\,(a)\;=\;{\rm tr}_{{\mathcal H}_\omega} \left( P\,\pi_\omega
  (a)\right), 
\label{eq:2.27}
\end{equation}
for all $a\in{\mathcal F}^r$. 

Let $\omega\; :=\; \omega_{\beta,\umu}$
be an infinite-volume equilibrium state on ${\mathcal F}^r$ obeying the
KMS condition (\ref{eq:2.21}). We assume that the Hilbert space
${\mathcal H}_\omega$ obtained from the GNS construction is {\it
  separable} and that the cyclic vector $\Omega\in{\mathcal H}_\omega$
is the {\it only eigenvector} (up to phases) of the operator
$L_\omega$ of (\ref{eq:2.26}), which, in this context, is called
the {\it Liouvillian} or {\it thermal Hamiltonian}. In other words,
the spectrum of $L_\omega$ is purely continuous, except for a simple
eigenvalue at 0.

This assumption implies the property of {\it ``return to
  equilibrium''}: If $\rho$ is an arbitrary state {\it normal}
relative to $\omega=\omega_{\beta,\umu}$ then
\begin{equation}
\lim_{T\to\infty} \frac 1 T \int_0^T {\rm dt}\,\rho
\left(\alpha_t(a)\right) \;=\; \omega(a),
\label{eq:2.28}
\end{equation}
for all $a\in{\mathcal F}^r$. If the spectrum of $L_\omega$ is {\it
  absolutely continuous}, except for a simple eigenvalue at 0, then
\begin{equation}
\lim_{t\to\infty} \rho\,\left(\alpha_t(a)\right)\;=\; \omega(a),
\label{eq:2.29}
\end{equation}
for all $a\in{\mathcal F}^r$. Eqs.~(\ref{eq:2.28}) and (\ref{eq:2.29})
follow from our assumptions on the spectrum of $L_\omega$
and the KMS condition (\ref{eq:2.21}); see e.g. [BFS].\\
\indent
Assuming the existence of the endomorphism $\sigma_+$, see (A4), we now
address the question of uniqueness and dynamical stability of the stationary
state $\omega_{\rm stat}$ in (\ref{eq:2.88}). \\
\indent
We suppose that the property of return to equilibrium, (\ref{eq:2.29}), holds for
each reservoir separately; (this can be shown for reservoirs consisting of
free fermions as considered in Sections 3 and 4). Recalling that the reference
state $\omega^0$ is a product of two KMS states (see (\ref{eq:2.69}),
(\ref{eq:2.66})), it is not difficult to extend the arguments in
Sect.~III,~D of [BFS] to show that if $\rho$ is an arbitrary state on
${\mathcal F}$ normal relative to the state $\omega^0$ in
(\ref{eq:2.69}) then
\begin{equation}
\lim_{t\to\infty}\;\rho \left(\alpha_t^0 (a)\right)\;=\;
\omega^0(a),~~a\in{\mathcal F}~, 
\label{eq:2.113}
\end{equation}
where $\alpha_t^0$ is the time evolution of the reservoirs before they
are coupled; see (\ref{eq:2.37}).

Eq.~(\ref{eq:2.113}) and the existence of a scattering endomorphism
$\sigma_+$, see eq.~(\ref{eq:2.89}), now imply that
\begin{eqnarray}
\lim_{t\to\infty}\;\rho_t(a) 
&=& \lim_{t\to\infty}\;\rho(\alpha_t(a))\nonumber\\
&=& \lim_{t\to\infty}\;\rho\bigl(\alpha_t^0\bigl(\sigma_+(a)
\bigr)\bigr)\nonumber\\
&=& \omega^0 \bigl(\sigma_+(a)\bigr)\nonumber\\
&=& \omega_{\rm stat} (a)~,
\label{eq:2.114}
\end{eqnarray}
for $a\in{\mathcal F}$. Thus, if the initial state $\rho$ is an {\it
  arbitrary} state {\it normal} relative to $\omega^0$ then the states
$\rho_t$ tend to the stationary state $\omega_{\rm stat}$, as
$t\to\infty$, ({\it uniqueness}).

Next, let $\rho$ be an {\it arbitrary} state {\it normal}
relative to $\omega_{\rm stat}$. We claim that if eq.~(\ref{eq:2.29})
holds for each reservoir then
\begin{equation}
\lim_{t\to\infty}\;\rho_t(a)\;=\;\omega_{\rm stat} (a),~ a\in{\mathcal
  F}.
\label{eq:2.115}
\end{equation}
Eq.~(\ref{eq:2.115}) is the property of {\it ``return to the
  stationary state'' (stability)}, for states $\rho$ {\it normal}
relative to $\omega_{\rm stat}$. To prove eq.~(\ref{eq:2.115}), we
follow the arguments in Sect.~III,~D of [BFS]: Since $\rho$ is normal
relative to $\omega_{\rm stat}$, there exist non-negative numbers,
$p_n, ~n=1,2,3,\ldots$, with $\sum_{n=1}^\infty p_n=1$, and nets of
operators $\{u_n^\alpha\}_{\alpha\in I_n}$, $n=1,2,3,\ldots$, with
$u_n^\alpha\in{\mathcal F}$, for all $\alpha$ and all $n$, such that
\[
\rho(a)\;=\sum_{n=1}^\infty p_n\;\lim_{\alpha,\alpha'}\;\big\langle \pi
\bigl( u_n^\alpha\bigr) \Omega, \pi(a)\,\pi\bigl(
  u_n^{\alpha'}\bigr)\Omega\big\rangle 
\]
where $\pi$ is the GNS representation and $\Omega$ the cyclic vector
corresponding to $(\omega_{\rm stat}, {\mathcal F})$; see
(\ref{eq:2.23}), (\ref{eq:2.24}), and (\ref{eq:2.27}). Then
\[
\rho_t(a)\;=\sum_{n=1}^\infty p_n\;\lim_{\alpha,\alpha'}\; \big\langle
\pi\bigl( u_n^\alpha\bigr) \Omega,\, \pi\bigl( \alpha_t(a)\bigr) \pi
\bigl( u_n^{\alpha'}\bigr)\Omega\big\rangle~.
\]
Let
\[
b_1\;:=\;u_n^\alpha~,~~ b_2\;:=\;u_n^{\alpha'}~,
\]
for some fixed $n,\alpha,\alpha'$. Then
\begin{equation}
\big\langle \pi (b_1)\Omega, \,\pi \bigl( \alpha_t(a)\bigr)\,\pi
(b_2)\Omega\big\rangle \;=\;\omega_{\rm stat} \bigl( b_1^* \alpha_t
(a) b_2\bigr)~. 
\label{eq:2.116}
\end{equation}
Since $\omega_{\rm stat} (a)\;=\; \omega^0 \bigl( \sigma_+(a)\bigr)~$,
and since, by eq.~(\ref{eq:2.89}),
\begin{equation}
\sigma_+\bigl(\alpha_t(a)\bigr)\;=\;\alpha_t^0\bigl(\sigma_+(a)\bigr)~,
\label{eq:2.117}
\end{equation}
for arbitrary $a\in{\mathcal F}$, the R.S. of (\ref{eq:2.116}) is 
given by
\[
\omega_{\rm stat}\bigl( b_1^*\alpha_t(a)b_2\bigr)\;=\; 
\omega^0\bigl(\sigma_+(b_1^*)\,\alpha_t^0\bigl(
\sigma_+(a)\bigr)\,\sigma_+(b_2)\bigr)~. 
\]
It follows from (\ref{eq:2.113}) by polarization that
\begin{eqnarray}
\lim_{t\to\infty}\; \omega^0\bigl(\sigma_+(b_1^*)\,
\alpha_t^0\bigl(\sigma_+(a)\bigr)\,\sigma_+(b_2)\bigr)
&=&\omega^0\bigl(\sigma_+(b_1^*)\,\sigma_+(b_2)\bigr)\,
\omega^0\bigl(\sigma_+(a)\bigr)\nonumber\\ 
&=&\omega_{\rm stat}\bigl( b_1^*\,b_2\bigr)\;\omega_{\rm stat}
(a)\nonumber\\
&=&\big\langle\pi(b_1)\Omega, \pi (b_2)\Omega\big\rangle\;
\omega_{\rm stat}(a)~.\phantom{mmmmm}
\label{eq:2.118}
\end{eqnarray}
Our contention, eq.~(\ref{eq:2.115}), follows from this.

\subsection{Cluster properties and profiles in ${\bf d}{\mathbf >}{\bf
    2}$ dimensional\\
  systems } \label{subsec-2.7} 

The last question we wish to address, in this summary of the general
theory, concerns {\it cluster properties} of the stationary state
$\omega_{\rm stat}$, which will show that $\omega_{\rm stat}$ {\it
  cannot} be an equilibrium (KMS) state for the dynamics, $\alpha_t$,
of the coupled reservoirs and is, in general, {\it not} normal
relative to the product state, $\omega^0$, of the uncoupled
reservoirs. 

We consider two increasing families of reservoirs
confined to regions $\Lambda^r\subset{\mathbb R}^3$, with
\begin{equation}
\Lambda^r\nearrow{\mathbb R}^3~,~~r\;=\;I,I\!I~,
\label{eq:2.119}
\end{equation}
joined together by a thermal contact or a tunnelling junction
localized near the origin, $\ux=0$, of physical space. The more
realistic situation where the reservoirs are confined to two
complementary half spaces, ${\mathbb R}_+^3$ and ${\mathbb
  R}_-^3$, respectively, with a junction localized near the origin, has
  been considered in [DFG]; see also [Ru1, Ru2]. It will be studied in
  more detail elsewhere. In order to describe spatial properties of the
  system, we make the following assumption.\\

{\bf (A5)~ Existence of space translations}.

{\it
For each reservoir $r=I, II$, there exists a $^*$automorphism (semi-) group
\begin{equation}
\left\{ \tau^r_{\ux}\bigm| \ux \in {\mathbb R}_{(\pm)}^3\right\}
\label{eq:2.30}
\end{equation}
of the field algebra ${\mathcal F}^r$, representing
{\it space translations} of ${\mathbb R}^3$ (${\mathbb R}_\pm^3$,
respectively) on ${\mathcal F}^r$.
}\\

\noindent
For the system of two coupled reservoirs,
\begin{equation}
\tau_\ux\;:=\;\tau_\ux^I\otimes\tau_\ux^{I\!I}~,~~\ux\in{\mathbb
  R}^3~, 
\label{eq:2.120}
\end{equation}
defines a representation of space translations as a 3-parameter group
of $^*$auto\-morph\-isms on the field algebra~ ${\mathcal F}=\overline{{\mathcal
  F}^I\otimes{\mathcal F}^{I\!I}}$. It
is plausible that space translations satisfy the following assumption.

\vspace{.5cm}

{\bf (A6)~ Asymptotic abelianness of space translations, and
  homogeneity\\
$\phantom{mmmml}$  of reservoirs}.

{\it
The action of $\tau_\ux$ on ${\mathcal F}$ is
norm-continuous in $\ux\in{\mathbb R}^3$ and for all operators
$a$ and $b$ in ${\mathcal F}$, 
\begin{equation}
\lim_{|\ux|\to\infty}\; \Vert \left[ \tau_\ux(a), b\right]\Vert\;=\;0~.
\label{eq:2.121}
\end{equation}
Furthermore, the dynamics and the
equilibrium states of the uncoupled reservoirs are {\it homogeneous}, in the
sense that 
\begin{equation}
\alpha_t^0\bigl(\tau_\ux(a)\bigr)\;=\;\tau_\ux\bigl(\alpha_t^0(a)\bigr)
\label{eq:2.122}
\end{equation}
and
\begin{equation}
\omega^0\bigl(\tau_\ux(a)\bigr)\;=\;\omega^0(a)~,
\label{eq:2.123}
\end{equation}
for all~ $a\in{\mathcal F}$. 
}\\

\noindent
The local nature of the perturbation, $W$, of the dynamics of the
system due to the contact or junction, see Assumption (A3),
eqs.~(\ref{eq:2.34}) and (\ref{eq:2.35}), and assumption (A6), then
imply that, for all $a\in{\mathcal F}$,
\begin{equation}
\lim_{|\ux|\to\infty}\; \Vert \alpha_t\bigl(\tau_\ux(a)\bigr) -
\alpha_t^0\bigl(\tau_\ux(a)\bigr)\Vert\;=\;0~, 
\label{eq:2.124}
\end{equation}
for all times $t$. A proof of (\ref{eq:2.124}) follows
from the Lie-Schwinger series for $\alpha_{-t}(\alpha_t^0(a))$ and use
of (\ref{eq:2.35}) and (\ref{eq:2.121}). Relation (\ref{eq:2.124}) shows that
observables localized far from the junction evolve according to the
non-interacting dynamics.\\
\indent
It is tempting, and can be justified in examples, to strengthen
assumption (A4) (existence of scattering endomorphism) as follows.\\

{\bf (A7)~ Cluster properties of the scattering endomorphism}.

{\it
The limits
\begin{equation}
\displaystyle\mathop{n-\lim}_{t\to\pm\infty}\;
\alpha_{-t}^0\bigl(\alpha_t\bigl(\tau_\ux(a)\bigr)\bigr)\;=\;
\sigma_\pm\bigl(\tau_\ux(a)\bigr) 
\label{eq:2.125}
\end{equation}
are {\it uniform} in $\ux\in{\mathbb R}^3$, for every $a\in{\mathcal
  F}$.
}\\

\noindent
Equations (\ref{eq:2.124}) and (\ref{eq:2.125}) imply that
\begin{equation}
\lim_{|\ux|\to\infty}\;\Vert\sigma_\pm\bigl(\tau_\ux(a)\bigr) -
\tau_\ux(a)\Vert\;=\;0~, 
\label{eq:2.126}
\end{equation}
for every $a\in{\mathcal F}$. From this property we conclude that 
\begin{eqnarray}
\lim_{|\ux|\to\infty}\;\omega_{\rm stat}\bigl(\tau_\ux(a)\bigr)
&=&\lim_{|\ux|\to\infty}\;\omega^0\bigl(\sigma_+
\bigl(\tau_\ux(a)\bigr)\bigr)\nonumber\\
&=& \lim_{|\ux|\to\infty}\;\omega^0\bigl(\tau_\ux(a)\bigr)\nonumber\\
&=&\omega^0(a),~~ a\in{\mathcal F}~,
\label{eq:2.127}
\end{eqnarray}
i.e., very far from the junction, the stationary state $\omega_{\rm
  stat}$ resembles the product state $\omega^0$ of the uncoupled
reservoirs.

\vspace{.5cm}

{\bf Remark}.~ It is not hard to understand that if the two reservoirs
occupy complementary half spaces, ${\mathbb R}_+^3$ and ${\mathbb
  R}_-^3$, then (\ref{eq:2.127}) is replaced by
\[
\lim_{x\to-\infty}\;\omega_{\rm stat}
\bigl(\tau_{(x,0,0)}\bigl(a\otimes\id\bigr)\bigr) \;=\;
\omega_{\beta^I,\umu^I}(a)~, 
\]
for $a\in{\mathcal F}^I$, and
\[
\lim_{x\to+\infty}\;\omega_{\rm stat}
\bigl(\tau_{(x,0,0)}\bigl(\id\otimes b\bigr)\bigr)\;=\;
\omega_{\beta^{I\!I},\umu^{I\!I}} (b)~,
\]
for $b\in{\mathcal F}^{I\!I}$. This may prove the presence of a {\it profile} of
temperature, density, or \dots in the stationary
state, $\omega_{\rm stat}$, of the system.

In the examples studied in subsequent sections, which concern junctions
between ordinary {\it three-dimensional} metals, assumptions (A6) and
(A7) can be verified. Instead of three-dimensional reservoirs, we
could consider {\it one-dimensional} (wires), or {\it two-dimensional}
(layers) reservoirs joined by a thermal contact or a tunnelling
junction. Our analysis in Sect.~4 will show that, in dimensions
$d=1,2$, assumption (A7) will {\it fail}, in general. In fact, the
example of the one-dimensional XY spin chain treated
in [DFG, AP] and the example of quantum wires studied in [ACF] show
that, in {\it one} dimension, $\omega_{\rm stat}$ may well be
space-translation invariant, i.e., it does not exhibit any profile. In
the example of quantum wires, $\omega_{\rm stat}$ is actually a {\it
  homogeneous thermal equilibrium state}. 

Thus, we observe that the validity of assumption (A7) critically
depends on the {\it dimension} of the reservoirs. In dimension $d>2$,
this assumption can be expected to hold, while it usually fails in
dimension $d=1,2$. For people familiar with elementary facts of
scattering theory this will not come
as a surprise.

In a subsequent paper, we will show that, for a large class of
reservoirs, one can construct {\it ``observables at infinity''}, see
e.g. [BR, Vol. II], corresponding to the operators $P^r(t)$ and $I_j^r(t)$
defined in eqs.~(\ref{eq:2.50}), (\ref{eq:2.56}),
respectively. Clearly, the expectation values of these operators {\it
  vanish} in the product state $\omega^0$ of the uncoupled reservoirs
and are given by ${\mathcal P}^r$ and ${\mathcal J}_j^r$ in the
stationary state, $\omega_{\rm stat}$, of the coupled reservoirs. If
we can show that ${\mathcal P}^r\neq 0$, or that ${\mathcal J}_j^r
\neq 0$, for $r=I$ or $I\!I$ and some $j$, then it follows that
$\omega_{\rm stat}$ is {\it not} normal relative to $\omega^0$. In the
examples studied in Sects.~4 and 5, we shall encounter instances where
${\mathcal P}^r$ and $\flow^r$ do {\it not} vanish.

\secct{Reservoirs of non-interacting fermions}
\label{sec-3}

This section serves to introduce a class of simple, but physically
important examples of reservoirs to which the general theory outlined
in Sect.~\ref{sec-2} can and will be applied. Our examples describe a
quantum liquid of non-interacting, non-relativistic electrons in a
normal metal or a semi-conductor, possibly subject to an external
magnetic field, or an ideal quantum gas of fermionic atoms or
molecules. In a subsequent paper, we shall also consider examples
describing chiral Luttinger liquids, which arise in connection with
the quantum Hall effect. 

We start by considering a system consisting of a single,
non-relativistic quant\-um-mechanical particle confined to a region
$\Lambda$ of physical space ${\mathbb R}^d$, $d=1,2,3$. The Hilbert
space of pure state vectors of this system is given by the space
\begin{equation}
L^2\bigl(\Lambda, d^dx\bigr)
\label{eq:3.1}
\end{equation}
of square-integrable wave functions with support in $\Lambda$. If the
particle has spin and/or if there are several species of such
particles then $L^2(\Lambda, d^dx)$ must be replaced by the space
\begin{equation}
h^\Lambda\;:=\;L^2\bigl(\Lambda, d^dx\bigr) \otimes {\mathbb C}^k~,
\label{eq:3.2}
\end{equation}
where $k=\sum_{\alpha=1}^l (2S_\alpha+1)$, $S_\alpha$ is the spin of
species $\alpha$, and $l$ is the number of species.

The one-particle dynamics is generated by the following selfadjoint operator,
$t^\Lambda$, acting on $h^\Lambda$,
\begin{equation}
t^\Lambda\;=\;- \frac{\hbar^2}{2M}~\Delta \otimes \id~,
\label{eq:3.3}
\end{equation}
where $M$ is the mass of the particle and $\Delta$ is the Laplace
operator on $L^2(\Lambda, d^dx)$ with selfadjoint boundary conditions (e.g. Dirichlet,
Neumann, or periodic) imposed at the boundary,
$\partial\Lambda$, of $\Lambda$.

In the following, we choose units in which $\hbar=1$ and $M=\frac 12$.

Other operators are physically interesting.
Electrons in semi-conductors would involve a potential operator that is diagonal in the space
representation. A magnetic field could also be considered; the Laplacian should be replaced by the
covariant Laplacian, and a coupling between the spin of the particle and the magnetic field should be
introduced. In this paper, we restrict our attention to the situation \eqref{eq:3.3}.

Next, we consider a system consisting 
of $n$ identical particles of the kind just considered, all confined
to the region $\Lambda$. Its state space is given by a subspace of the
$n$-fold tensor product of $h^\Lambda$ of fixed symmetry type,
\begin{equation}
h_n^\Lambda \;:=\;P(h^\Lambda)^{\otimes n},~~
h_0^\Lambda\;:=\;{\mathbb C}~,
\label{eq:3.6}
\end{equation}
where $P$ is the orthogonal projection onto the subspace of wave
functions of the selected symmetry type under permutations of the $n$
particle variables. If the particles are {\it bosons} then $P\equiv
P_+$ projects onto {\it completely symmetric} $n$-particle wave
functions; while, for {\it fermions}, $P\equiv P_-$ projects onto {\it
  totally anti-symmetric} wave functions. In this paper, we focus our
attention on fermions.

If the particles do not interact with each other the Hamiltonian,
$T_n^\Lambda$ of the $n$-particle system is given by 
\begin{equation}
T_n^\Lambda \;:=\;\sum_{j=1}^n \id\otimes\ldots\otimes t_j^\Lambda
\otimes\ldots\otimes \id~,
\label{eq:3.7}
\end{equation}
where $t_j^\Lambda$ acts on the $j^{\rm th}$ factor in the $n$-fold
tensor product in (\ref{eq:3.6}).

If the number of particles can fluctuate (e.g., because the system is
coupled to a particle reservoir such as a battery) then it is
convenient to use the formalism of ``second quantization'', which we
briefly recall.

The {\it Fock space} is defined by
\begin{equation}
{\mathcal H}^\Lambda\;:=\; \bigoplus_{n=0}^\infty\; h_n^\Lambda~.
\label{eq:3.8}
\end{equation}
The free dynamics on ${\mathcal H}^\Lambda$ is generated by the {\it
  Hamiltonian}
\begin{equation}
H^\Lambda\;:=\;\bigoplus_{n=0}^\infty\; T_n^\Lambda~,
\label{eq:3.9}
\end{equation}
with $T_n^\Lambda$ as in (\ref{eq:3.7}). The {\it particle number
  operator}, $N^\Lambda$, is defined by
\begin{equation}
N^\Lambda \;:=\;\bigoplus_{n=0}^\infty\; n\cdot\id\big|_{h_n^\Lambda}~.
\label{eq:3.10}
\end{equation}
Let $\kappa$ be a symmetric $k\times k$ matrix acting on ${\mathbb 
  C}^k$. We set
\[
K_n^\Lambda\;:=\;\sum_{j=1}^n \id \otimes\ldots\otimes \bigl(
\id\otimes\kappa_j\bigr) \otimes\ldots\otimes\id~, 
\]
where $\id\otimes\kappa_j$ acts on the $j^{\rm th}$ factor, $\simeq L^2
(\Lambda, d^dx)\otimes{\mathbb C}^k$, in the $n$-fold tensor product
defining $h_n^\Lambda$. For $t^\Lambda$ as in (\ref{eq:3.3}), a
typical charge operator, $Q^\Lambda\equiv Q^\Lambda(\kappa)$, is of
the form
\begin{equation}
Q^\Lambda\;:=\;\bigoplus_{n=0}^\infty\; K_n^\Lambda~.
\label{eq:3.11}
\end{equation}

The operators $H^\Lambda, N^\Lambda$ and $Q^\Lambda(\kappa)$ are
unbounded, selfadjoint operators on ${\mathcal H}^\Lambda$; see
e.g. [RS].

Next, we describe the structure of ${\mathcal H}^\Lambda$ in some more
detail and introduce creation and annihilation operators. Let
$\ux,\uy,\ldots$ denote points in physical space ${\mathbb R}^d$, and
let $s=1,\ldots,k$ label
on orthonormal basis in ${\mathbb C}^k$. Vectors $f_n$ in the
$n$-particle space $h_n^\Lambda$ can be represented as
square-integrable wave functions,
\[
f_n\left(\ux_1,s_1,\ldots,\ux_n,s_n\right)~,
\]
with support in $\Lambda^{dn}\subset{\mathbb R}^{dn}$, which, for
fermions, are totally anti-symmetric under permutations of their $n$
arguments. Vectors $\psi,\phi,\ldots$ in Fock space correspond to
sequences,
\begin{equation}
\psi\;=\;(f_n)_{n=0}^\infty~~,~~\phi\;=\;(g_n)_{n=0}^\infty~,\ldots 
\label{eq:3.12}
\end{equation}
of $n$-particle wave functions in $h_n^\Lambda$. The scalar product
on ${\mathcal H}^\Lambda$ is defined by
\begin{equation}
\langle\psi,\phi\rangle\;:=\sum_{n=0}^\infty \sum_{s_1,\ldots,s_n} 
\int_\Lambda \prod_{j=1}^nd\ux_j \overline{f_n(\ux_1,s_1,\ldots,\ux_n,s_n)}
\,g_n(\ux_1,s_1,\ldots,\ux_n,s_n)~.
\label{eq:3.13}
\end{equation}
The vector represented by the sequence $(f_n)_{n=0}^\infty$, with
$f_0=1$, $f_n\equiv 0$, for $n\geq 1$, is denoted by $\Omega$
and is called the {\it vacuum (vector)}.

Let ${\mathcal D}={\mathcal D}^\Lambda$ be the linear domain of
vectors $\psi=(f_n)_{n=0}^\infty$ in ${\mathcal H}^\Lambda$ with the
property that {\it all} but {\it finitely many} $f_n$'s {\it
  vanish}. Clearly, ${\mathcal D}$ is dense in ${\mathcal
  H}^\Lambda$. For $f\in h^\Lambda$, we define an {\it annihilation
  operator}, $a(f)$, by
\begin{eqnarray}
&&\bigl(a(f)\psi\bigr)_n\bigl(\ux_1,s_1,\ldots,\ux_n,s_n\bigr)\::=
\nonumber\\
&&~~~~~~~~\sqrt{n+1} \sum_{s=1}^k \int_\Lambda d
\ux\;\overline{f(\ux,s)}\, f_{n+1}\bigl(
\ux,s,\ux_1,s_1,\ldots,\ux_n,s_n\bigr)~, 
\label{eq:3.14}
\end{eqnarray}
for arbitrary $\psi=(f_n)_{n=1}^\infty \in{\mathcal D}$, and
\begin{equation}
a(f)\Omega\;:=\;0~.
\label{eq:3.15}
\end{equation}
The {\it creation operator}, $a^*(f)$, is defined to be the adjoint of
$a(f)$ on ${\mathcal H}^\Lambda$ and is easily seen to be well defined
on ${\mathcal D}$.

It is well known, see e.g. [RS, BR], that, for {\it fermions}, the
following {\it ``canonical anti-commutation relations''} (CAR) hold:
\begin{equation}
\left\{ a^\#(f),~a^\#(g)\right\}\;=\;0~,
\label{eq:3.16}
\end{equation}
for arbitrary $f,g$ in $h^\Lambda$ where $a^\#=a$ or $a^*$, and
$\{A,B\}:= AB+BA$ is the anti-commutator of two operators $A$ and $B$;
\begin{equation}
\left\{ a(f),~a^*(g)\right\}\;=\;(f,g)\cdot\id~,
\label{eq:3.17}
\end{equation}
where $(f,g):=\sum_s\int_\Lambda d\ux\,\overline{f(\ux,s)}\,g(\ux,s)$
is the scalar product on $h^\Lambda$. For bosons, (\ref{eq:3.16}) and
(\ref{eq:3.17}) hold if anti-commutators are replaced by commutators
(CCR). Formally,
\[
a(f)\;=\sum_s\int_\Lambda d\ux\;\overline{f(\ux,s)}\;a(\ux,s),
\]
and
\[
a^*(f)\;=\sum_s\int_\Lambda d\ux\;a^*(\ux,s)\, f(\ux,s)~,
\]
with
\begin{equation}
\left\{ a(\ux,s),\,a^*(\ux',s')\right\} \;=\;\delta_{ss'}\,
\delta^{(d)} (\ux-\ux')~.
\label{eq:3.18}
\end{equation}

A remarkable consequence of the CAR is
that the operators $a(f)$ and $a^*(f)$ are {\it bounded} in norm by 
\begin{equation}
\Vert a(f)\Vert\;=\;\Vert a^*(f)\Vert\;=\;\Vert f\Vert\;:=\;
\sqrt{(f,f)}~. 
\label{eq:3.19}
\end{equation}
To see this, we choose an arbitrary $\psi\in{\mathcal D}$ and note
that
\begin{eqnarray*}
&& \Vert a(f)\psi\Vert^2\,+\,\Vert a^*(f)\psi\Vert^2\\
&&\phantom{mmm} =\;\big\langle a(f)\psi,\,a(f)\psi\big\rangle\,+\,
\big\langle a^*(f)\psi,\,a^*(f)\psi\big\rangle\\
&&\phantom{mmm} =\; \big\langle \psi, \left\{ a(f),\,a^*(f)\right\}
  \psi\big\rangle\\
&&\phantom{mmm} =\;(f,f)\,\langle\psi,\psi\rangle~,
\end{eqnarray*}
so that
\begin{equation}
\Vert a^\# (f)\psi\Vert \;\leq\;\Vert f\Vert \cdot\Vert\psi\Vert~.
\label{eq:3.20}
\end{equation}
Equality in (\ref{eq:3.20}) is seen from examples.

Eq.~(\ref{eq:3.19}) is false for {\it bosons}, $a(f)$ and $a^*(f)$
being {\it unbounded} operators. 

For fermions, polynomials in $a(f)$, $a^*(f)$, $f\in h^\Lambda$, form a
$^*$algebra of operators on ${\mathcal H}^\Lambda$ 
which is weakly dense in $B({\mathcal H}^\Lambda)$. The ``observable
algebra'' ${\mathcal A}^\Lambda$ is the norm closure of the algebra
of these polynomials in $a(f), a^*(f)$, $f\in h^\Lambda$, which
commute with the number operator $N^\Lambda$ and, possibly, with
further charge operators $Q^\Lambda(\kappa)$, for certain choices of
$\kappa$. Every monomial in $a$ and $a^*$ belonging to ${\mathcal
  A}^\Lambda$ has equally many factors of $a$ and $a^*$, since it must
conserve the total particle number. A general monomial in $a$ and
$a^*$ is {\it Wick-ordered} if all $a^*$'s are to the left of all
$a$'s.

In terms of creation and annihilation operators, the operators
$H^\Lambda, N^\Lambda$ and $Q^\Lambda$ can be expressed as follows.
\begin{equation}
H^\Lambda\;=\sum_s\int_\Lambda d\ux\, a^*(\ux,s)(t^\Lambda a)(\ux,s)~,
\label{eq:3.21}
\end{equation}
\begin{equation}
N^\Lambda\;=\sum_s\int_\Lambda d\ux\,a^*(\ux,s)\,a(\ux,s)~,
\label{eq:3.22}
\end{equation}
and
\begin{equation}
Q^\Lambda(\kappa)\;=\sum_{s,s'}\int_\Lambda d\ux\, a^*(\ux,s)\,
\kappa_{ss'}\,a(\ux,s')~. 
\label{eq:3.23}
\end{equation}

In the examples discussed below and in Sects.~4 and 5, we usually
regard $N^\Lambda=Q^\Lambda(\kappa=\id)$ to be the {\it only
  conservation law}, besides $H^\Lambda$, relevant for the description
of the reservoirs. In a general
discussion, we consider $M$ conservation laws, $Q_j^\Lambda=Q^\Lambda
(\kappa_j)$, $j=1,\ldots,M$, and choose $t^\Lambda$ as in (\ref{eq:3.3}).

The main result of this section is the following theorem. 


\begin{theorem} \label{thm-3.1}
  For $t^\Lambda$ as in (\ref{eq:3.3}), and
  $Q_j^\Lambda=Q_j^\Lambda (\kappa_j), j=1,\ldots,M$, with
  $\kappa_1,\ldots,\kappa_M$ arbitrary, commuting symmetric $k\times
  k$ matrices, the equilibrium states $\omega_{\beta,\umu}^\Lambda$
  introduced in eq.~(\ref{eq:2.10}) exist, for arbitrary $\beta \geq
  0$ and $\umu \in {\mathbb R}^M$. 

Assumptions (A1), (A2), (A5) and (A6) of Sect.~2, concerning the
existence of the thermodynamic limit, $\Lambda\nearrow{\mathbb R}^d$, hold.
\end{theorem}

The proof of Theorem~\ref{thm-3.1} is standard. A careful exposition
can be found in [BR], Sect.~5.2.

In Section~4 we shall consider a system consisting of two identical
reservoirs, $I$ and $I\!I$, both composed of non-interacting,
non-relativistic fermions confined to some region 
$\Lambda=\Lambda^I=\Lambda^{I\!I}$ of ${\mathbb R}^d$. A convenient
notation for creation and annihilation operators for the two
reservoirs is the following one.
\begin{eqnarray}
a^\#(\ux,s,I) &:=& a^\#(\ux,s)\big|_{{\mathcal H}^{\Lambda^I}} \otimes
\id\big|_{{\mathcal H}^{\Lambda^{I\!I}}}~,
\nonumber\\
a^\#(\ux,s,I\!I) &:=& \id\big|_{{\mathcal H}^{\Lambda^I}} \otimes a^\#
(\ux,s)\big|_{{\mathcal H}^{\Lambda^{I\!I}}}~.
\label{eq:3.24}
\end{eqnarray}
We note that all the operators $a^\#(f,I)=a^\#(f)\otimes \id$ {\it
  commute} with all the operators $a^\#(f,I\!I)=\id\otimes
a^\#(f)$. If, for convenience, we prefer that they anti-commute we can
accomplish this feature by a standard {\it Klein-Jordan-Wigner
  transformation}:
\begin{eqnarray}
a^\#(f,I) &\mapsto& a^\#(f,I)~,\nonumber\\
a^\#(f,I\!I) &\mapsto& a^\#(f,I\!I)\;
e^{i\pi(N^{\Lambda^I}\otimes\id)}~. 
\label{eq:3.25}
\end{eqnarray}
The operators on the R.S. of (\ref{eq:3.25}) will again be denoted by
$a^\#(f,r)$, $r=I,I\!I$. We introduce the following notation. 
\begin{eqnarray}
X &:=& (\ux,s,r)\in{\mathbb R}^d \times \{ 1,\ldots,k\} \times \{
I,I\!I\}, \label{eq:3.26}\\
X^{(N)}&:=&\left( X_1,\ldots,X_N\right)~,\label{eq:3.28}\\
x^{(N)}&:=& (\ux_1, \ldots,\ux_n)\in{\mathbb R}^{dN},\\
s^{(N)}&:=& (s_1,\ldots,s_N)\in \{1,\ldots,k\}^N,\\
r^{(N)}&:=& (r_1,\ldots,r_N)\in \{I, II\}^N,\\
\int_\Lambda dX  &:=& \sum_{r=I,I\!I}\;\sum_{s=1}^k
\int_\Lambda d\ux, \label{eq:3.27}\\
\int_{\Lambda^N} d X^{(N)} &:=& \prod_{j=1}^N \int_\Lambda
d X_j ,\label{eq:3.29}\\
{\mathbf a}^\# (X^{(N)}) &:=& \prod_{j=1}^N a^\#(X_j)~,
\label{eq:3.30}
\end{eqnarray}
with $a^\#=a^*$ or $a$.

We are now prepared to describe the interactions,
$W(\Lambda^I,\Lambda^{I\!I})$, (see eq.~{\ref{eq:2.34})),
  corresponding to thermal contacts or tunnelling junctions between
  the two reservoirs. We shall always assume that the {\it total}
  particle number of the system consisting of the two reservoirs is
  conserved. Thus the interaction Hamiltonian $W(\Lambda^I, \Lambda^{I\!I})$ must commute with the operator
\begin{equation}
N^{I\cup I\!I} \;:=\;N^{\Lambda^I} \otimes \id + \id \otimes
N^{\Lambda^{I\!I}}~, 
\label{eq:3.31}
\end{equation}
(see eq.~(\ref{eq:2.43})). It follows that $W(\Lambda^I,
\Lambda^{I\!I})$ must have the form 
\begin{equation}
W(\Lambda^I,\Lambda^{I\!I}) \;=\sum_{N=1}^\infty
W_N(\Lambda^I,\Lambda^{I\!I})~, 
\label{eq:3.32}
\end{equation}
where
\begin{equation}
W_N(\Lambda^I,\Lambda^{I\!I})\;=\int_{\Lambda^N} d X^{(N)}
\int_{\Lambda^N} d Y^{(N)} {\mathbf a}^*(X^{(N)})\,
w_N^{\Lambda^I,\Lambda^{I\!I}} (X^{(N)},Y^{(N)})\, {\mathbf
    a}(Y^{(N)})~, 
\label{eq:3.33}
\end{equation}
and, for each choice of $s_1,r_1,\ldots$, $s_N,r_N$,
$s_1',r_1',\ldots$, $s_N',r_N'$,
\[
w_N^{\Lambda^I,\Lambda^{I\!I}}\bigl(\bigl(x^{(N)},s^{(N)},r^{(N)}\bigr),
\bigl(y^{(N)},{s'}^{(N)},{r'}^{(N)}\bigr)\bigr)
\]
is a smooth function of $x^{(N)}\in\Lambda^N$ and
$y^{(N)}\in\Lambda^N$ vanishing if $x^{(N)}\not\in \Lambda^N$ or
$y^{(N)}\not\in\Lambda^N$. In Sect.~4, we introduce weighted Sobolev
  spaces, ${\mathcal W}_N$ equipped with norms $\Vert|(\cdot)\Vert|_N$ with
  the property that
\begin{equation}
\Vert W_N (\Lambda^I,\Lambda^{I\!I})\Vert\;\leq\;|\Vert
w_N^{\Lambda^I,\Lambda^{I\!I}}\Vert|_{N}~, 
\label{eq:3.34}
\end{equation}
for all $N=1,2,3,\ldots$. We shall assume that, for each $N$, there is
a function $w_N\in{\mathcal W}_N$ such that 
\begin{equation}
g(w)\;:=\;\sum_{N=1}^\infty \Vert| w_N\Vert|_N\,<\,\infty~, 
\label{eq:3.35}
\end{equation}
where $w=(w_N)_{N=1}^\infty$, and
\begin{equation}
\lim_{\substack{\Lambda^I\nearrow\infty \\ \Lambda^{I\!I}\nearrow\infty}}\;
\sum_{N=1}^\infty \Vert| w_N^{\Lambda^I,\Lambda^{I\!I}} -
w_N\Vert|_N\;=\;0~. 
\label{eq:3.36}
\end{equation}
It then follows that
\begin{equation}
\displaystyle\mathop{n-\lim}_{\substack{\Lambda^I\nearrow\infty \\
  \Lambda^{I\!I}\nearrow\infty}}\; W(\Lambda^I,\Lambda^{I\!I})\;=:\;W
\label{eq:3.37}
\end{equation}
exists, and
\begin{equation}
\Vert W\Vert\;\leq\;g (w)~.
\label{eq:3.38}
\end{equation}
Thus, assumption (A4), eq.~(\ref{eq:2.36}), of Sect.~2 follows from
(\ref{eq:3.33}), (\ref{eq:3.34}), (\ref{eq:3.35}) and (\ref{eq:3.36}). 

If we want to describe {\it thermal contacts} we shall require that
\begin{equation}
[N^{\Lambda^I}\otimes\id,\;W(\Lambda^I,\Lambda^{I\!I})]\;=\;[\id\otimes
N^{\Lambda^{I\!I}}, W(\Lambda^I,\Lambda^{I\!I})]\;=\;0~,
\label{eq:3.39}
\end{equation}
while, for {\it tunnelling junctions}, only
\begin{equation}
[ N^{I\cup I\!I}, W(\Lambda^I,\Lambda^{I\!I})]\;=\;0
\label{eq:3.40}
\end{equation}
is required, for arbitrary
$\Lambda^I=\Lambda^{I\!I}=\Lambda\subset{\mathbb R}^d$.

To conclude this section, we remark that a system of two reservoirs of
non-interacting fermions, with a one-particle Hamiltonian 
$t^\Lambda$ as in  Eq.~(\ref{eq:3.3}), and with interactions 
$W(\Lambda^I,\Lambda^{I\!I})$ as in (\ref{eq:3.33})-(\ref{eq:3.37}),
satisfies Assumptions (A1)-(A3) and (A5), (A6) of Section 2. Assumptions (A4)
and (A7) are established in the next section for $d\geq 3$ and small $g(w)$.



\secct{Existence of M\o ller endomorphisms} \label{sec-4}

The goal of this section is to illustrate the general theory of Section
\ref{sec-2} by providing a complete 
mathematical description of a concrete system, namely two coupled reservoirs
of free fermions in dimension $d\geq3$. An illustration can be found in Fig.\ \ref{figreservoirs}.
The reservoirs are infinite and without boundary, and the coupling $W$ is
localized near the origin, in
the the sense that $\|W\|'$, or $\|W\|''$, is finite (see \eqref{normW}, \eqref{defnorm}, or
\eqref{notanorm}).

\begin{figure}[htb]
\epsfxsize=80mm
\centerline{\epsffile{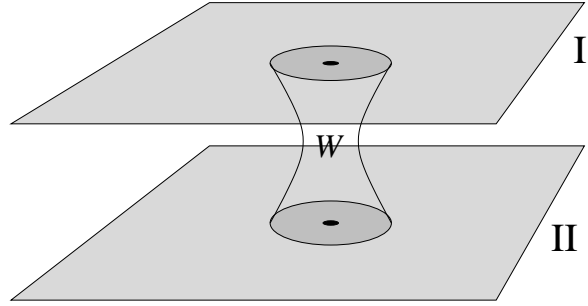}}
\caption{Two-dimensional reservoirs coupled by a local interaction $W$. We actually consider the
three-dimensional analogue of this situation.}
\label{figreservoirs}
\end{figure}

The Hamiltonian for each reservoir has been introduced in Section \ref{sec-3}, see
equation (\ref{eq:3.9}). The coupling between
reservoirs is represented by the interaction in equations (\ref{eq:3.32}) and (\ref{eq:3.33}). As stated in Theorem \ref{thm-3.1},
there exist time evolution automorphisms $\alpha_t^0$ and $\alpha_t$ in the
thermodynamic limit. The former corresponds to the free dynamics and the latter to the dynamics for interacting
reservoirs.

In this section, we establish the existence of the M\o ller endomorphisms $\sigma_\pm$
defined in (\ref{eq:2.89}). We start with the Dyson series for $\alpha_{-t}^0 \alpha_t$, namely
\begin{equation}
\label{Dyson}
\alpha_{-t}^0 \alpha_t(a) = \sum_{m=0}^\infty i^m \int_{t>t_m>\dots>t_1>0} d t_1 \dots d t_m \; [W(t_m), \dots,
[W(t_1),a] \dots ],
\end{equation}
where we set $W(t):= \alpha_{-t}^0(W)$. Convergence of this series for finite
$t$ is clear since 
$\|W\|<\infty$, but we need to consider the limit $t\to\infty$. Let us define
an operator $D_m(t)$ on the field algebra ${\cal F}$ by
\begin{equation}
\label{defDm}
D_m(t) a := \int_{t>t_m>\dots>t_1>0} d t_1 \dots d t_m \; [W(t_m), \dots,
[W(t_1),a] \dots ],
\end{equation}
for arbitrary $a\in{\cal F}$. It is understood that $D_0 a = a$.

We define the norm of an interaction $W$ by setting
\begin{multline}
\label{normW}
\|W\|' = \sum_{N\geq1} N 2^{(d+2)N} \sum_{s_1, \dots, s_N = 1}^k \sum_{s_1', \dots, s_N' = 1}^k
\\
\sum_{r_1, \dots, r_N = I,II} \sum_{r_1', \dots, r_N' = I,II} \|
w_N \bigl( (\cdot, s^{(N)}, r^{(N)}), (\cdot, {s'}^{(N)}, {r'}^{(N)}) \bigr) \|'_{2dN};
\end{multline}
the function $w_N$ is viewed above as a function on ${\mathbb R}^{2dN}$, and
the norm $\|\cdot\|_M'$ is 
defined by
\begin{equation}
\label{defnorm}
\|f\|_M' = \frac1{2^{3M/2}} \biggl[ \int_{{\mathbb R}^M} d x^{(M)} \overline{f(x^{(M)})}
\prod_{k=1}^M \Bigl( -\frac{d^2}{d x_k^2} + x_k^2 + 1 \Bigr)^3 f(x^{(M)}) \biggr]^{1/2}.
\end{equation}
Note that the operators $-\frac{d^2}{d x_k^2} + x_k^2 + 1$ are bounded below
by 2, which 
implies the inequality (\ref{eq:3.34}), namely $\|f\|_{L^2({\mathbb R}^M)}
\leq \|f\|'_M$; this inequality is
saturated when $f$ is a product of Gaussians centered at the origin,
$f(x^{(M)}) = \prod_{k=1}^M e^{-x_k^2/2}$.

\begin{theorem}
\label{thmDyson}
For $d\geq3$ we have the bound
$$
\int_0^\infty \| [W(t), D_{m-1}(t) a] \| d t \leq \frac1m \Bigl( \frac{8\pi d}{d-2} \Bigr)^m \|a\|'
\; {\left(\|W\|'\right)}^m.
$$
\end{theorem}
\ \\

A similar statement holds for negative times. That is, one can rewrite the Dyson series in
(\ref{Dyson}) and the operator $D_M(t)$ in (\ref{defDm}) for $t<0$ by integrating over negative
times $0>t_1>\dots>t_m>t$. Then Theorem \ref{thmDyson} holds with an integral from $-\infty$ to
0. The proof for negative times is identical to the one of positive times.

Before proving Theorem \ref{thmDyson}, let us work out its main consequence, the existence of M\o
ller endomorphisms.

\begin{corollary}
\label{corMoller}
If $\frac{8\pi d}{d-2} \|W\|'<1$, there exist $\sigma_\pm$ such that
$$
\lim_{t\to\pm\infty} \| \alpha_{-t}^0 \alpha_t(a) - \sigma_\pm(a) \| = 0
$$
for all $a$ with $\|a\|'<\infty$.
\end{corollary}

Corollary \ref{corMoller} implies the existence of a scattering automorphism, see Assumption (A4).
We comment below that the norm \eqref{defnorm} can be replaced by an object that is translation
invariant, see \eqref{notanorm}. Therefore the scattering automorphism is given by a limit of
infinite times, and this limit exists in norm, uniformly with respect
to space translations (for both the interaction and the operator $a\in\mathcal
F$). Hence Assumption (A7) holds.

\begin{proof}[Proof of Corollary \ref{corMoller}]
Observe that
\begin{equation}
\alpha_{-t}^0 \alpha_t(a) = a + \sum_{m\geq1} i^m \int_0^t [W(s), D_{m-1}(s) a] d s.
\end{equation}
Then, for $t<t'$,
\begin{equation*}
\| \alpha_{-t}^0 \alpha_t(a) - \alpha_{-t'}^0 \alpha_{t'}(a) \| \leq \sum_{m\geq1} \int_t^{t'} \|
[W(s), D_{m-1}(s) a] \| d s.
\end{equation*}
By Theorem \ref{thmDyson} and the dominated convergence theorem, the right side vanishes as $t,t' \to
\infty$. This implies the norm-convergence of $\alpha_{-t}^0 \alpha_t(a)$.
\end{proof}

We will make use of Hermite functions in the proof of Theorem \ref{thmDyson};
so we collect a few useful facts on them. 
 The Hermite functions are denoted by $\{\phi_q\}_{q\in{\mathbb N}}$, where
\begin{equation*}
\phi_q(x) = \frac{(-1)^q}{\sqrt{2^q q!} \pi^{\frac14}} e^{\frac12 x^2}
\Bigl( \frac d{d x}
\Bigr)^q e^{-x^2},
\end{equation*}
with $x\in{\mathbb R}$. These functions satisfy the equation
\begin{equation}
\label{diffeqHermite}
\Bigl( -\frac{d^2}{d x^2} + x^2 \Bigr) \phi_q(x) = (2q+1) \phi_q(x).
\end{equation}

\begin{lemma}\hfill
\label{lemHermite}

\begin{itemize}
\item[(i)] $\| \phi_q \|_1 \leq \sqrt{4\pi (q+1)}$.
\item[(ii)] $ |(e^{i t \Delta} \phi_p,\phi_q)| \leq \frac{\|\phi_p\|_1
\|\phi_q\|_1}{\sqrt{4\pi|t|}}$.
\end{itemize}
\end{lemma}

\begin{proof}
By Cauchy-Schwarz and since the operator $-\frac{d^2}{d x^2}$ is positive
definite, we have that
\begin{align}
\|\phi_q\|_1 &= \int_{-\infty}^\infty \sqrt{x^2+1} \, |\phi_q(x)| \, \frac1{\sqrt{x^2+1}} d x
\nonumber\\
&\leq \Bigl( \phi_q, \Bigl( -\frac{d^2}{d x^2} + x^2 + 1 \Bigr) \phi_q \Bigr)^{1/2} \Bigl( \int_{-\infty}^\infty
\frac{d x}{x^2+1} \Bigr)^{1/2}.
\end{align}
The first factor is equal to $\sqrt{2(q+1)}$ and the second one equals 
$\sqrt{2\pi}$, which proves (i). Claim (ii) immediately follows from
\begin{equation*}
\Bigl( e^{i t \Delta} \phi_p, \phi_q \Bigr) = \sqrt{\frac i{4\pi t}} \int_{-\infty}^\infty d x
\int_{-\infty}^\infty d y e^{-i (x-y)^2 /4t} \phi_p(y) \phi_q(x).
\end{equation*}
\end{proof}

Hermite functions form an orthonormal basis of $L^2({\mathbb
  R})$. We use them to express the interaction as a
polynomial of creation and annihilation operators of fermions in states
  described by Hermite functions. The free 
time evolution of the interaction can be described as an evolution of these
  functions, and their decorrelation in 
time can be controlled using Lemma \ref{lemHermite} (ii).  Finally, Hermite functions will be removed
at the expense of introducing differential operators in the definition of the
  norm of the interaction; see \eqref{defnorm}.

We use from now on the following notation: for $\uq=(q_1,\ldots,q_d) \in {\mathbb N}^d$ and $\ux=(x_1,\ldots,x_d)
\in {\mathbb R}^d$,
\begin{equation}
\phi_\uq(\ux) = \prod_{i=1}^d \phi_{q_i}(x_i).
\end{equation}

\begin{proof}[Proof of Theorem \ref{thmDyson}]
We start by rewriting the interaction in the basis of Hermite functions. Let
$Q^{(N)}:=(Q_1,\ldots,Q_N)$, with 
\begin{equation}
Q_j = (\uq_j, s_j, r_j) \in {\mathbb N}^d \times \{1,\dots,k\} \times \{\rm{I,II}\}.
\end{equation}
We set 
\begin{equation}
\label{wtilde}
\tilde w_N(Q^{(N)}, {Q'}^{(N)}) = \int_{{\mathbb R}^{dN}} d x^{(N)}
\int_{{\mathbb R}^{dN}} d y^{(N)}
w_N(X^{(N)}, Y^{(N)}) \prod_{j=1}^N \phi_{\uq_j}(\ux_j) \phi_{\uq_j'}(\uy_j).
\end{equation}
Here, $X^{(N)}$ is determined by $Q^{(N)}$ and $x^{(N)}$, namely $X_j = (\ux_j, s_j, r_j)$ if
$Q_j = (\uq_j, s_j, r_j)$. The interaction (\ref{eq:3.31}),
(\ref{eq:3.32}) is given by a sum over 
\begin{equation*}
\uQ:=(N, Q^{(N)},{Q'}^{(N)}), 
\end{equation*}
\begin{equation}
W = \sum_\uQ W_\uQ = \sum_\uQ {\mathbf a}^*(Q^{(N)}) \tilde w_N(Q^{(N)}, {Q'}^{(N)})
{\mathbf a}({Q'}^{(N)}),
\end{equation}
where ${\mathbf a}^*(Q^{(N)}) = \prod_{j=1}^N a^*(Q_j)$,  and $a^*(Q_j)$ is the
creation operator for a fermion in the state $\phi_{\uq_j}$, of spin $s_j$, in
the reservoir $r_j$. The annihilation operators ${\mathbf a}({Q'}^{(N)})$
are defined similarly. The operator $a$ also has a Hermite expansion 
\begin{equation}
a = \sum_\uQ a_\uQ = \sum_\uQ {\mathbf a}^*(Q^{(N)}) \tilde a_N(Q^{(N)}, {Q'}^{(N)})
{\mathbf a}({Q'}^{(N)}).
\end{equation}

With this notation, we have that 
\begin{multline}
\label{expansion}
\int_0^\infty \| [W(t),D_{m-1}(t) a] \| d t \leq \sum_{\uQ_0, \dots, \uQ_m}
\int_{\infty>t_m>\dots>t_1>0} d t_1
\dots d t_m \\
\| [W_{\uQ_m}(t_m), \dots, [W_{\uQ_1}(t_1),a_{\uQ_0}] \dots ] \|.
\end{multline}

The multiple commutator above involves operators $W$ and $a$, which in turn involve creation and
annihilation operators of particles in both reservoirs. The latter satisfy anticommutation relations
for particles in the same reservoir, or commutation relations for particles in different reservoirs.
This introduces a complication when estimating the multicommutator above. This complication can be
avoided by using the Klein-Jordan-Wigner transformation explained in (\ref{eq:3.25}). For simplicity, we keep
the same notation, but we assume from now on that all creation and annihilation operators satisfy
anticommutation relations.

Because $\alpha_{-t}^0$ is a $*$-automorphism, its action on the interaction
$W$ simply amounts to replacing
operators $a^\#(Q) = a^\#(\phi_\uq,s,r)$ by
\begin{equation}
\alpha_{-t}^0(a^\#(\phi_\uq,s,r)) = a^\#(e^{i t \Delta} \phi_\uq, s, r) := a^\#(Q,t).
\end{equation}
We note that Lemma \ref{lemHermite} yields the bound
\begin{align}
\label{timestimate}
\left\| \left\{ \alpha_{-t}^0(a^\#(Q)), \alpha_{-t'}^0(a^\#(Q')) \right\} \right\| &= |(e^{i t \Delta} \phi_\uq,
e^{i t' \Delta} \phi_{\uq'})| \delta_{ss'} \delta_{rr'} \nonumber\\
&\leq \bigl( 1 \wedge \tfrac{4\pi}{|t-t'|} \bigr)^{d/2} \prod_{i=1}^d (q_i+1)^{\frac12}
(q_i'+1)^{\frac12}.
\end{align}

The multicommutator in \eqref{expansion} can be written as
\begin{multline}
\label{multicommutator}
[W_{\uQ_m}(t_m), \dots, [W_{\uQ_1}(t_1), a_{\uQ_0}] \dots] = \\
\tilde
a_{N_0}(Q_0^{(N_0)}, {Q_0'}^{(N_0)}) \prod_{j=1}^m \tilde w_{N_j}(Q_j^{(N_j)},
{Q_j'}^{(N_j)}) 
\big[{\mathbf a}^*(Q_m^{(N_m)},t_m) {\mathbf a}({Q_m'}^{(N_m)},t_m), \dots\\
\dots,
\big[{\mathbf a}^*(Q_1^{(N_1)},t_1) {\mathbf a}({Q_1'}^{(N_1)},t_1),
{\mathbf a}^*(Q_0^{(N_0)})
{\mathbf a}({Q_0'}^{(N_0)})\big] \dots\big].
\end{multline}
Here, we set
\begin{equation}
{\mathbf a}^\#(Q_j^{(N_j)},t_j) = \prod_{\ell=1}^{N_j}
\alpha_{-t_j}^0({\mathbf a}^\#(Q_{j,\ell})),
\end{equation}
where $Q_{j,\ell}$ is the $\ell$-th element of $Q_j^{(N_j)}$.

A commutator of products of operators can be expanded according to contraction
schemes. 
The following equation holds when $k\ell$ is even:
\begin{multline}
[a_1 \dots a_k, b_1 \dots b_\ell] = \sumtwo{1\leq i\leq k}{1\leq j\leq\ell} (-1)^{i\ell+j+1} a_1 \dots a_{i-1} b_1
\dots b_{j-1} \\
\{a_i,b_j\} b_{j+1} \dots b_\ell a_{i+1} \dots a_k.
\end{multline}
The multicommutator of \eqref{multicommutator} can thus be expanded in
contraction schemes for operators
at different times. An operator at time $t_1$ contracts necessarily with an operator at time
$t_0=0$; an operator at time $t_2$ contracts with an operator at time $t_{r_2}$ with $r_2 = 0,1$;
\dots; an operator at time $t_m$ contracts with an operator at time $t_{r_m}$ with $r_m =
0,\dots,m-1$. See Fig.\ \ref{figtree} for an illustration.
\begin{figure}[htb]
\epsfxsize=120mm
\centerline{\epsffile{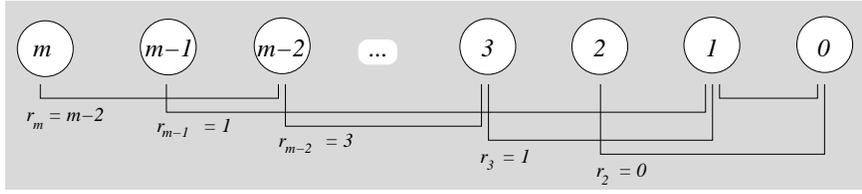}}
\caption{Illustration for the numbers $r_m,\dots,r_2$ that occur in the choice of contractions. We
see that they define a tree.}
\label{figtree}
\end{figure}
To a set of contractions corresponds a monomial of creation and annihilation
operators, multiplied by anticommutators of contracted operators.

The monomial of creation and annihilation operators is bounded in operator
norm by 1. Contracted operators are
estimated using \eqref{timestimate}. This yields a factor involving times,
namely 
$$
\prod_{j=1}^m \bigl( 1 \wedge \tfrac{4\pi}{t_j-t_{r_j}} \bigr)^{d/2}.
$$
Second, one obtains a factor involving indices of Hermite functions for the
contracted operators. An upper bound on this factor is obtained by writing a product over all indices, namely
$$
\prod_{j=0}^m \prod_{k=1}^{N_j} \prod_{i=1}^d (q_{jki}+1)^{\frac12} (q_{jki}'+1)^{\frac12}.
$$
Here, $\uq_{jk}, \uq_{jk}' \in {\mathbb N}^d$ are indices for Hermite functions determined by the $k$-th element of
$\uQ_j$.

It remains to estimate the number of contraction schemes, given
$r_2,\dots,r_m$. We define 
\begin{equation}
\label{incidence}
e_j = \bigl| \{k:r_k=j\} \bigr| + 1 - \delta_{j0}, \quad 0\leq j\leq m.
\end{equation}
Notice that $1\leq e_0 \leq m$ and $1\leq e_j \leq m-j+1$ if $j\neq0$. $e_j$ is the
number of operators at time $t_j$ that belong to a contraction and it is necessarily smaller than
$2N_j$. Since there are $2N_j$ operators
at time $t_j$, the number of possible contractions is
$$
\prod_{j=0}^m \frac{(2N_j)!}{(2N_j-e_j)!}.
$$
The above estimates could be improved by observing that many contraction
schemes yield zero; namely, in the case where both
operators are creation or annihilation operators; or if the spins are different; or if they belong to different
reservoirs. It is not easy to take advantage of this, however.

We now gather the above estimates to obtain the bound
\begin{multline}
\int_0^\infty \| [W(t), D_{m-1}(t) A] \| d t \leq \\
\sum_{\uQ_0, \dots, \uQ_m} |\tilde
A_{N_0}(Q_0^{(N_0)},{Q_0'}^{(N_0)})| \prod_{j=1}^m |\tilde
w_{N_j}(Q_j^{(N_j)},{Q_j'}^{(N_j)})| \\
\times \prod_{j=0}^m \prod_{k=1}^{N_j} \prod_{i=1}^d
(q_{jki}+1)^{\frac12} (q_{jki}'+1)^{\frac12} \int_{\infty>t_m>\dots>t_1>0} d t_1 \dots d t_m \\
\times \sum_{r_m=0}^{m-1} \sum_{r_{m-1}=0}^{m-2} \dots \sum_{r_2=0}^1 \prod_{j=1}^m \bigl( 1 \wedge
\tfrac{4\pi}{t_j-t_{r_j}} \bigr)^{d/2} \prod_{j=0}^m \frac{\chi[e_j \leq 2N_j] (2N_j)!}{(2N_j-e_j)!}.
\end{multline}

A sequence of numbers $r_2, \dots, r_m$ can be represented by a graph with set of vertices
$\{0,\dots,m\}$, and an edge between $i$ and $j$ whenever $r_j=i$. This graph is a tree: there are $m$ edges, and each vertex
$j\neq0$ is directly connected to a vertex $i<j$, hence each vertex is eventually connected to 0. The
numbers $e_j$ defined in (\ref{incidence}) are then the incidence numbers of the tree --- $e_j$ is the
number of edges containing the vertex $j$. This is illustrated in Fig.\ \ref{figtree}. We can
symmetrize the bound by summing over {\it all}
trees $T$ with $m+1$ vertices; this step will allow to deal with the time integrals. Reorganizing, we obtain
\begin{multline}
\int_0^\infty \| [W(t), D_{m-1}(t) a] \| d t \leq \\
\sum_{\uQ_0} |\tilde
a_{N_0}(Q_0^{(N_0)},{Q_0'}^{(N_0)})| \prod_{k=1}^{N_0} \prod_{i=1}^d (q_{0ki}+1)^{\frac12}
(q_{0ki}'+1)^{\frac12} \\
\times \int_{\infty>t_m>\dots>t_1>0} d t_1 \dots d t_m \sum_T \prod_{(i,j)\in T} \bigl(
1 \wedge \tfrac{4\pi}{|t_i-t_j|} \bigr)^{d/2} \frac{\chi[e_0 \leq 2N_0] (2N_0)!}{(2N_0-e_0)!} \\
\times \prod_{j=1}^m \bigg\{ \sum_{\uQ_j} |\tilde w_{N_j}(Q_j^{(N_j)},{Q_j'}^{(N_j)})|
\prod_{k=1}^{N_j} \prod_{i=1}^d (q_{jki}+1)^{\frac12} (q_{jki}'+1)^{\frac12} \frac{\chi[e_j \leq
2N_j] (2N_j)!}{(2N_j-e_j)!} \biggr\}.
\end{multline}
Let us focus on the time integrals. The integrand is a symmetric function of $t_1,\dots,t_m$, because
of the sum over arbitrary trees. We can
therefore extract a factor $1/m!$, at the cost of integrating over {\it all} positive times $t_1,\dots,t_m$ without the
ordering condition. Since there is no integral over $t_0=0$, we have 
\begin{equation}
\int_{0}^\infty d t_1 \dots \int_{0}^\infty d t_m \prod_{(i,j)\in T} \bigl( 1 \wedge
\tfrac{4\pi}{|t_i-t_j|} \bigr)^{d/2} \leq \Bigl( \int_{-\infty}^\infty \bigl( 1 \wedge
\tfrac{4\pi}{|t|} \bigr)^{d/2} d t \Bigr)^m,
\end{equation}
for any tree $T$. 
The last integral is equal to $8\pi d/(d-2)$.

The number of trees with $m+1$ vertices  and incidence numbers $e_0, \dots, e_m$ is equal to
$$
\Bigl( \begin{matrix} m-1 \\ \; e_0-1, \; e_1-1, \; \dots, \; e_m-1 \; \end{matrix} \Bigr) =
\frac{(m-1)!}{(e_0-1)! \dots (e_m-1)!};
$$
see for instance [Ber], Th\'eor\`eme 2 p.\ 86. We sum over incidence numbers, using $\sum_{e=0}^{2N} (\begin{smallmatrix} 2N
\\ e \end{smallmatrix} ) = 4^N$, and we get
\begin{multline}
\int_0^\infty \| [W(t), D_{m-1}(t) a] \| d t \leq \frac1m \Bigl( \frac{8\pi d}{d-2} \Bigr)^m
\sum_{\uQ_0} N_0 4^{N_0} |\tilde a_{N_0}(Q_0^{(N_0)},{Q_0'}^{(N_0)})| \\
\times\prod_{k=1}^{N_0}
\prod_{i=1}^d (q_{0ki}+1)^{\frac12} (q_{0ki}'+1)^{\frac12}\\
\times \bigg\{ \sum_{\uQ} N 4^N |\tilde
w_N(Q^{(N)},{Q'}^{(N)})| \prod_{k=1}^N \prod_{i=1}^d (q_{ki}+1)^{\frac12}
(q_{ki}'+1)^{\frac12} \biggr\}^m.
\end{multline}

The last step consists in removing the Hermite functions. We fix $N$, $s^{(N)}$, ${s'}^{(N)}$,
$r^{(N)}$, ${r'}^{(N)}$, and perform the summation over $q^{(N)}$ and ${q'}^{(N)}$. Using Cauchy-Schwarz, we obtain
\begin{multline}
\label{stillvalid}
\sum_{q^{(N)}, {q'}^{(N)}} \left|\tilde w_N(Q^{(N)},{Q'}^{(N)})\right| \prod_{k=1}^N \prod_{i=1}^d
(q_{ki}+1)^{\frac12} (q_{ki}'+1)^{\frac12} \\
\leq \biggl( \sum_{q^{(N)}, {q'}^{(N)}} \left|\tilde w_N(Q^{(N)},{Q'}^{(N)})\right|^2 \prod_{k=1}^N
\prod_{i=1}^d (q_{ki}+1)^3 (q_{ki}'+1)^3 \biggr)^{1/2}\\
\times \biggl( \sum_{\nu\geq0} \frac1{(q+1)^2}
\biggr)^{dN}.
\end{multline}
The last factor on the right side is bounded by $2^{dN}$. The first
factor on the right side can be viewed as an expectation value of a certain
operator 
expressed in the basis of Hermite functions. Rewriting it in the $x$-space
representation, we find that it is given by the square root of the following
expression 
\begin{multline}
\frac1{2^{6dN}} \int_{{\mathbb R}^{dN}} d x^{(N)} \int_{{\mathbb R}^{dN}} d
y^{(N)}\ 
\overline{w_N(X^{(N)},Y^{(N)})} \\
\times \prod_{k=1}^N \prod_{i=1}^d \Bigl( -\frac{d^2}{d x_{ki}^2} + x_{ki}^2 + 1 \Bigr)^3 \Bigl(
-\frac{d^2}{d y_{ki}^2} + y_{ki}^2 + 1 \Bigr)^3 w_N(X^{(N)},Y^{(N)}). \nonumber
\end{multline}
This motivates the use of the norm \eqref{defnorm} and concludes the proof of Theorem \ref{thmDyson}.
\end{proof}

We end this section by remarking that an estimate can be obtained that is invariant under space
translations. Such an estimate follows by repeating the steps above with translates of Hermite functions.
Recall that $\tilde w_N$ was defined in \eqref{wtilde} by integrating $2dN$ Hermite functions
centered at the origin. We can choose $z\in\mathbb R^{2dN}$ and translate the $j$-th function by $z_j$.
Lemma \ref{lemHermite} still holds with translates of Hermite functions, so
that the proof goes through
 without a change until \eqref{stillvalid}. Since a Hermite function translated by
$z\in\mathbb R$
satisfies the differential equation \eqref{diffeqHermite} with $(x-z)^2$ instead of $x^2$, one
gets a bound where the differential operators in the norm \eqref{defnorm} are translated by
$z\in\mathbb R^M$. This holds for all $z$; let us introduce $\|\cdot\|_M''$ by
\begin{eqnarray}
\label{notanorm}
\lefteqn{\|f\|_M'' =}\nonumber \\
&&\frac1{2^{3M/2}} \inf_{y \in \mathbb R^M} \biggl[ \int_{\mathbb R^M} d x^{(M)} \overline{f(x^{(M)})}
\prod_{k=1}^M \Bigl( -\frac{d^2}{d x_k^2} + (x_k-y_k)^2 + 1 \Bigr)^3 f(x^{(M)}) \biggr]^{1/2}.\nonumber\\
\end{eqnarray}
This object is translation invariant but it is not a norm. We have
$||\cdot||''_M\leq ||\cdot||'_M$. Theorem \ref{thmDyson} holds when $\|A\|'$
and $\|W\|'$ are replaced by $\|A\|''$ and $\|W\|''$, whose definition is like
\eqref{normW} with $\|\cdot\|_M''$ instead of $\|\cdot\|_M'$.

\secct{Explicit perturbative calculation of particle and energy currents}
\label{sec-5} 

In this section, we consider two reservoirs of non-relativistic
non-interacting free spinless fermions. Such systems are a special case of the
ones introduced in Section \ref{sec-3}. \\
\indent
For explicit calculations, it is convenient to represent the system in Fourier
space, see Subsection \ref{subsec-5.1}, since the one-particle energy
operator $t=-\Delta$ is diagonal in this representation. 
Subsection \ref{subsec-5.2} is devoted to the calculation of the particle and
energy currents 
for tunnelling junctions, in the lowest non-vanishing order in $W$. This 
establishes the relation between the particle current and chemical potentials
of 
the reservoirs, the {\it current voltage characteristics}. If the difference
of chemical potentials $\Delta\mu=\mu^I-\mu^{II}$ is small, then the particle
current is proportional to the voltage drop $\Delta\mu$. This linear relation
is known as {\it Ohm's law}. We calculate the
(inverse of the) proportionality factor, which is called the {\it resistance}
of the junction. Moreover, we explicitly
verify that the entropy production rate is strictly positive, provided the two
reservoirs are at either different temperatures or chemical potentials. \\

Let us recall that the single particle Hilbert space (in the
thermodynamic limit) is $h=L^2({\mathbb R}^d, d\ux)$, with $\ux\in {\mathbb
  R^d}, d\geq 3$ (see
equation (\ref{eq:3.2})). The dynamics is determined by $t=-\Delta$, see
(\ref{eq:3.3}). For each reservoir, we take the particle number to be the only
conservation law. Recall that for {\it tunnelling junctions}, the interaction
$W$ commutes with the total particle number operator, $N\otimes\id +\id\otimes
N$, while for {\it thermal junctions}, $W$ commutes separately with
$N\otimes\id$ and $\id\otimes N$; see equations (\ref{eq:3.40}) and
(\ref{eq:3.39}).\\
\indent
To quantify the interaction, we introduce two coupling constants, $g$ and $\xi$, and set
\begin{equation}
\label{wgxi}
W=g\sum_{N=1}^\infty \xi^N W_N.
\end{equation}
Let ${\cal J}_{k,l}, {\cal P}_{k,l}$ denote the term of order $g^k\xi^l$ of
the particle current $\cal J$ (see (\ref{eq:2.92}), (\ref{eq:2.99})
(\ref{eq:2.87})) and the energy current $\cal P$ (see (\ref{eq:2.91}),
(\ref{eq:2.98}), (\ref{eq:2.88})). Accordingly, we define ${\cal E}_{k,l}$,
where ${\cal E}$ is the entropy production rate in (\ref{eq:2.78}) and
(\ref{eq:2.84}).  
We give now explicit expressions for some lower order terms of the currents
and the entropy production rate. The calculations are presented in Subsection
5.2.\\

{\bf Tunnelling junctions.\ }
The lowest order terms of the particle current are
given by 
\begin{eqnarray}
{\cal J}_{1,1}&=&{\cal J}_{1,2}=0,\label{eq:5.10.1}\\
{\cal J}_{2,2}&=&2\pi\int_{{\mathbb R}^{2d}} d \uk\ d\ul\  \delta(|\uk|^2-|\ul|^2)
\left|\widehat{w}_1((-\uk,II),
  (\ul,I))\right|^2\left(\rho_{II}(\uk)-\rho_I(\uk)\right),\nonumber\\
&&\label{eq:5.10}
\end{eqnarray}
where $\uk, \ul\in {\mathbb R}^d$, $\widehat{w}$ is the Fourier transform of $w$, and the function $\rho_r(\uk)$
is defined as
\begin{equation*}
\rho_r(\uk)=\frac{1}{e^{\beta^r(|\uk|^2-\mu^r)}+1}.
\end{equation*}
We obtain for the energy current the expressions
\begin{eqnarray}
{\cal P}_{1,1}&=&{\cal P}_{1,2}=0,\label{eq:5.14.1}\\
{\cal P}_{2,2}&=&2\pi\int_{{\mathbb R}^{2d}} d\uk\ d\ul\
|\uk|^2\delta(|\uk|^2-|\ul|^2) 
\left|\widehat{w}_1((-\uk,II),
  (\ul,I))\right|^2\left(\rho_{II}(\uk)-\rho_I(\uk)\right).\nonumber\\
&& \label{eq:5.14}
\end{eqnarray}
Assuming that $\widehat{w}_1$ is not identically zero, the above formulas show
the following qualitative behaviour of the flows. 
\begin{enumerate}
\item[-] If $(\beta^I,\mu^I)=(\beta^{II},\mu^{II})$ then ${\cal J}_{2,2}={\cal
    P}_{2,2}=0$. The
   flows vanish if both reservoirs are at the same temperature and
  chemical potential.
\item[-] If $\mu^I=\mu^{II}$ and $\beta^I>\beta^{II}$ then
  $\rho_{II}(k)-\rho_I(k)>0$, for all $k$. Consequently, ${\cal J}_{2,2}, {\cal P}_{2,2}>0$. At
  constant chemical potential, there is a particle and energy flow from the
  hotter to the colder reservoir.
\item[-] If $\mu^I>\mu^{II}$ and $\beta^I=\beta^{II}=\beta$, then
    $\rho_{II}(k)-\rho_I(k)<0$, for all $k$. Consequently, ${\cal J}_{2,2}, {\cal P}_{2,2}<0$. At
    constant temperature, there is a particle and energy flow from the
    reservoir with higher chemical potential to the reservoir with lower
    chemical potential. 
\end{enumerate}

{\it Ohm's law and the resistance of the junction.\ }
Suppose that $\beta^I=\beta^{II}$ and $\mu^I=\mu, \mu^{II}=\mu+\Delta\mu$, with $\Delta\mu$ 
small. Retaining only the leading order in $\Delta \mu$ in the expression of
the particle flow yields
\begin{equation}
{\cal J}_{2,2}\approx\frac{\Delta\mu}{R(\mu,\beta)},
\label{eq:5.11}
\end{equation}
where the {\it resistance} $R(\mu,\beta)$ is determined by
\begin{equation}
R(\mu,\beta)^{-1} =2\pi\beta\int_{{\mathbb R}^{2d}}d\uk\ d\ul\
\delta(|\uk|^2-|\ul|^2) 
\frac{|\widehat{w}_1((-\uk,II),(\ul,I))|^2\
  e^{\beta(|\uk|^2-\mu)}}{(e^{\beta(|\uk|^2-\mu)}+1)^2}.
\label{resistance}
\end{equation}
We refer to Subsection 5.2 for a qualitative discussion of the resistance, in
three dimensions, $d=3$. \\

{\it Onsager reciprocity relations.} Let us study the interdependence of the flows near equilibrium. The
relevant parameters are the difference of the inverse temperatures, and the difference of the
chemical potentials divided by the temperature. Precisely, we set $\beta^I = \beta$; $\beta^{II} =
\beta - \Delta\beta$; $\nu = \beta^I \mu^I$; $\Delta\nu = \beta^I \mu^I - \beta^{II} \mu^{II}$. We
consider the flows to depend on $\beta$, $\nu$, $\Delta\beta$, and $\Delta\nu$.

One easily checks that
\begin{align}
&\frac\partial{\partial\Delta\beta} \bigl[ \rho_{II}(\uk) - \rho_I(\uk) \bigr] \Big|_{\Delta\beta =
\Delta\nu = 0} =
\frac{e^{\beta(|\uk|^2-\mu)}}{(e^{\beta(|\uk|^2-\mu)} + 1)^2} |\uk|^2, \nonumber\\
&\frac\partial{\partial\Delta\nu} \bigl[ \rho_{II}(\uk) - \rho_I(\uk) \bigr] \Big|_{\Delta\beta =
\Delta\nu = 0} = -
\frac{e^{\beta(|\uk|^2-\mu)}}{(e^{\beta(|\uk|^2-\mu)} + 1)^2}.
\end{align}
The first partial derivative is taken at constant $\beta$, $\nu$, and $\Delta\nu$; the second
partial derivative is at constant $\beta$, $\nu$, and $\Delta\beta$.
Then from \eqref{eq:5.10} and \eqref{eq:5.14} we observe that
\begin{equation}
\frac{\partial{\cal P}_{2,2}}{\partial\Delta\nu} \Big|_{\Delta\beta = \Delta\nu = 0} =
- \frac{\partial{\cal J}_{2,2}}{\partial\Delta\beta} \Big|_{\Delta\beta = \Delta\nu = 0}.
\end{equation}
This is an Onsager reciprocity relation and we see that it holds at lowest order. \\

{\it Entropy production rate.\ }
 Recall that
${\cal P}^{II}=-{\cal P}^I$ (equation (\ref{eq:2.98})) and ${\cal
  J}^{II}=-{\cal J}^I$ (equation (\ref{eq:2.99})), hence 
\begin{equation*}
{\cal E}=(\beta^I-\beta^{II}){\cal P} -(\beta^I\mu^I-\beta^{II}\mu^{II}){\cal
  J}.
\end{equation*}
Using the above expressions for ${\cal J}_{k,l}$ and ${\cal P}_{k,l}$, we
obtain 
\begin{eqnarray*}
{\cal E}_{1,1}&=&{\cal E}_{1,2}=0,\\
{\cal E}_{2,2}&=&(\beta^I-\beta^{II}){\cal P}_{2,2}
-(\beta^I\mu^I-\beta^{II}\mu^{II}){\cal J}_{2,2}\\
&=&2\pi\int_{{\mathbb R}^{2d}} d\uk\ d\ul \ \delta(|\uk|^2-|\ul|^2)
|\widehat{w}_1((-\uk,II),(\ul,I))|^2\\
&&\times\frac{\{(\beta^I-\beta^{II})|\uk|^2-(\beta^I\mu^I-\beta^{II}\mu^{II})\} \{
  e^{\beta^I(|\uk|^2-\mu^I)} -
  e^{\beta^{II}(|\uk|^2-\mu^{II})}\}}{(e^{\beta^I(|\uk|^2-\mu^I)}+1)(e^{\beta^{II}(|\uk|^2-\mu^{II})}+1)}.
\end{eqnarray*}
The numerator of the fraction is of the form
$(x^I-x^{II})(e^{x^I}-e^{x^{II}})$, with $x^r=\beta^r(|\uk|^2-\mu^r)$, hence it
is strictly positive unless $x^I=x^{II}$. We assume that
\begin{equation}
\int_{{\mathbb R}^d}d\ul\ \delta(|\uk|^2-|\ul|^2) |\widehat{w}_1((-\uk,II),(\ul,I))|^2
\label{eq:5.15}
\end{equation}
does not vanish for all $\uk\in{\mathbb R}^d$. Then ${\cal E}_{2,2}$ is strictly
positive unless $\beta^I(|\uk|^2-\mu^I)=\beta^{II}(|\uk|^2-\mu^{II})$ for all $\uk$
in the support of (\ref{eq:5.15}). This shows that {\it ${\cal E}_{2,2}$ is
  strictly positive unless $(\beta^I,\mu^I)=(\beta^{II},\mu^{II})$, in which
  case ${\cal E}_{2,2}$ vanishes.}\\

{\bf Thermal junctions.\ } The particle current is zero, a thermal junction
allows only for an exchange of heat between the two reservoirs. Since $W_1=0$,
the lowest order term which is nonvanishing is ${\cal P}_{2,4}$. Without
loss of generality, we take the coupling function $\widehat{w}_2$ to be of the
form 
\begin{equation*}
\widehat{w}_2((\uk_1,r_1), (\uk_2, r_2); (\ul_1,s_1), (\ul_2, s_2))
=\delta_{r_1, I}\ \delta_{r_2, II}\ \delta_{s_1, I}\ \delta_{s_2, II}\  \widehat{w}_2(\uk_1, \uk_2, \ul_1, \ul_2).
\end{equation*}
A somewhat lengthy but straightforward calculation yields
\begin{eqnarray}
\lefteqn{
{\cal P}_{2,4} =2\pi\int d\uk_1 d\uk_2 d\ul_1 d\ul_2 \ \delta( |\uk_1|^2+|\uk_2|^2-|\ul_1|^2-|\ul_2|^2)} \nonumber\\
&& \times |\widehat{w}_2(-\uk_1,-\uk_2,\ul_1,\ul_2)|^2 
(|\uk_1|^2 -|\ul_1|^2)\rho_I(\ul_1)\rho_{II}(\ul_2) (1-\rho_I(\uk_1) -\rho_{II}(\uk_2)),\nonumber\\ 
\label{m2}
\end{eqnarray}
from which we obtain the following qualitative discussion.
\begin{enumerate}
\item[-] If $(\beta^I, \mu^I)=(\beta^{II}, \mu^{II})$, then $\rho_I=\rho_{II}$, and by switching $\ul_1\leftrightarrow \ul_2$, $\uk_1\leftrightarrow \uk_2$
in the integral, we see that ${\cal P}_{2,4}=0$.
\item[-]
By splitting the integral in (\ref{m2}) into a sum of two integrals over the
regions $\chi(|\uk_1|^2>|\ul_1|^2)$ and $\chi(|\uk_1|^2<|\ul_1|^2)$, and switching $\uk_1\leftrightarrow \ul_1$, $\uk_2\leftrightarrow \ul_2$, we can rewrite 
\begin{eqnarray*}
\lefteqn{
{\cal P}_{2,4} =2\pi\int d\uk_1 d\uk_2 d\ul_1 d\ul_2 \ \delta( |\uk_1|^2+|\uk_2|^2-|\ul_1|^2-|\ul_2|^2)} \nonumber\\
&& \times |\widehat{w}_2(-\uk_1,-\uk_2,\ul_1,\ul_2)|^2 \ 
(|\uk_1|^2 -|\ul_1|^2)\ \chi(|\uk_1|^2>|\ul_1|^2)\\
&&\times  \big\{\rho_I(\ul_1)\rho_{II}(\ul_2) [1-\rho_I(\uk_1) -\rho_{II}(\uk_2)]\\
&&\ \ \ \ \ \ 
 - \rho_I(\ul_1)\rho_{II}(\ul_2) [1-\rho_I(\uk_1) -\rho_{II}(\uk_2)]\big\}\\
&=&2\pi\int d\uk_1 d\uk_2 d\ul_1 d\ul_2 \ \delta( |\uk_1|^2+|\uk_2|^2-|\ul_1|^2-|\ul_2|^2) \nonumber\\
&& \times |\widehat{w}_2(-\uk_1,-\uk_2,\ul_1,\ul_2)|^2 \ 
(|\uk_1|^2 -|\ul_1|^2)\ \chi(|\uk_1|^2>|\ul_1|^2)\\
&& \times   \big\{\rho_{II}(\ul_2) [1
-\rho_{II}(\uk_2)][\rho_I(\ul_1)-\rho_I(\uk_1)]\\
&&\ \ \ \ \ \ - \rho_I(\uk_1)[1-\rho_I(\ul_1)][\rho_{II}(\uk_2)-\rho_{II}(\ul_2)]\big\}.
\end{eqnarray*}
The first product in the round brackets $\{\ \}$ is strictly positive and
tends to zero, as $\beta^{II}\rightarrow\infty$ (because in the limit
$\beta^r\rightarrow\infty$, 
$\rho_r(k)$ tends to the characteristic function $\chi(|k|^2<\mu^r)$). The
second term in the round brackets is strictly negative and tends to zero, as
$\beta^{I}\rightarrow\infty$. We conclude that ${\cal P}_{2,4}<0$ if
$\beta^I<\infty$, and $\beta^{II}$ is large enough; as expected!
\end{enumerate}

\subsection{Fourier representation} \label{subsec-5.1}
The creation and annihilation operators in the Fourier representation are
defined by
\begin{eqnarray}
a^*(\uk,r)&=&(2\pi)^{-d/2}\int_{{\mathbb R}^d} d\ux \ e^{i\uk\ux} a^*(\ux,r),\label{eq:5.1}\\ 
a(\uk,r)&=&(2\pi)^{-d/2}\int_{{\mathbb R}^d} d\ux\  e^{-i\uk\ux} a(\ux,r),\label{eq:5.2}
\end{eqnarray}
where $\uk\in{\mathbb R}^d$, $r=I,II$; compare with (\ref{eq:3.18}) and
(\ref{eq:3.24}). The dynamics of $a^*(\uk,r)$ and $a(\uk,r)$ is given by
\begin{equation}
\alpha_t^r(a^*(\uk,r))=e^{i\omega t}a^*(\uk,r),\ \ \alpha_t^r(a(\uk,r))=e^{-i\omega
  t} a(\uk,r),
\label{eq:5.3}
\end{equation}
where 
\begin{equation*}
\omega=\omega(\uk)=|\uk|^2.
\end{equation*}
The operators $H^r, N^r, W_N$ defined in (\ref{eq:3.21}), (\ref{eq:3.22}),
(\ref{eq:3.33}) are represented in Fourier space (and in the thermodynamic
limit) by
\begin{eqnarray}
H^r&=&\int_{{\mathbb R}^d} d\uk\ \omega(\uk) a^*(\uk,r)a(\uk,r), 
\label{eq:5.4}\\
N^r&=&\int_{{\mathbb R}^d} d\uk\ a^*(\uk,r)a(\uk,r),\nonumber\\
W_N&=&\int d K^{(N)} d L^{(N)} {\mathbf a}^*(K^{(N)})
\widehat{w}_N(-K^{(N)},L^{(N)}) {\mathbf a}(L^{(N)}),
\nonumber
\end{eqnarray}
where we introduce notation analogous to (\ref{eq:3.26})-(\ref{eq:3.30}). For
$K^{(N)}=(K_1,\ldots,K_N)$, $K_j=(\uk_j,r_j)\in{\mathbb
  R}^d\times \{I,II\}$, we set $-K^{(N)}:=(-K_1,\ldots,-K_N)$,
where $-K_j=(-\uk_j,r_j)$. The symbol \ $\widehat{\ }$ \ denotes the Fourier
transform, i.e.
\begin{eqnarray*}
\lefteqn{
\widehat{w}_N(K^{(N)},L^{(N)})}\\
&=&(2\pi)^{-dN}
\int_{{\mathbb R}^{2dN}}d\ux_1\cdots
d\uy_Ne^{-i(\uk_1\ux_1+\cdots +\ul_N\uy_N)}
w_N(X^{(N)},Y^{(N)}).
\end{eqnarray*}
We recall some properties of the state $\omega^0$ defined in
(\ref{eq:2.69}), which is given by
\begin{equation}
\omega^0= \omega_{\beta^I,\mu^I}\otimes \omega_{\beta^{II},\mu^{II}},
\label{eq:5.5}
\end{equation}
where $\omega_{\beta^r,\mu^r}$ is the equilibrium state of reservoir $r=I,II$
in the thermodynamic limit; see also Theorem \ref{thm-3.1}. The two-point
function of $\omega_{\beta^r,\mu^r}$ is
\begin{equation*}
\omega_{\beta^r, \mu^r}(a^*(\uk,r)a(\ul,r))=\delta(\uk-\ul)\rho_r(\uk),\mbox{\ \
  where } \ \ 
\rho_r(\uk)=\frac{1}{e^{\beta^r(|\uk|^2-\mu^r)}+1}.
\end{equation*}
The average of a monomial in $n$ creation and $m$ annihilation operators is
zero unless 
$n=m$, in which case it can be calculated recursively from the formula
\begin{eqnarray*}
\lefteqn{\omega_{\beta^r,\mu^r}\left(\prod_{i=1}^n a^*(\uk_i,r) \prod_{j=1}^n
 a(\ul_j,r)\right)}\\
&&=\sum_{p=1}^n (-1)^{n-p} \omega_{\beta^r,\mu^r}(a^*(\uk_1,r)a(\ul_p,r)) 
\ \omega_{\beta^r,\mu^r}\left( \prod_{i=2}^n a^*(\uk_i,r)\prod_{j=1, j\neq p}^n
 a(\ul_j,r)\right).
\end{eqnarray*}
For details, we refer to [BR]. We are now ready for explicit
calculations of the currents.

\subsection{Calculations for tunnelling junctions} \label{subsec-5.2}

{\bf Particle current and resistance.\ } The particle current 
\begin{equation}
{\cal J}=\omega_{\rm stat}(-i[N\otimes \id,W])=\omega^0(\sigma_+(-i[N\otimes
\id,W]))
\label{eq:5.6}
\end{equation}
has been introduced in (\ref{eq:2.92}), (\ref{eq:2.99}), see also
(\ref{eq:2.87}). We set $\hbar =1$. We see from the Dyson series expansion of
$\sigma_+$, see (\ref{Dyson}), and the definition of ${\cal J}_{k,l}$ (see
after (\ref{wgxi})), that
\begin{eqnarray}
{\cal J}_{1,1}&=& -i\omega^0([N\otimes\id, W_1]),\nonumber\\
{\cal J}_{1,2}&=& -i\omega^0([N\otimes\id, W_2]), \nonumber\\
{\cal J}_{2,2}&=&\int_0^\infty dt\ 
\omega^0([W_1(t),[N\otimes\id, W_1]]).
\label{eq:5.7}
\end{eqnarray}
It is not difficult to verify that
\begin{eqnarray}
[N\otimes\id, W_N]&=&\int d K^{(N)}
d L^{(N)}\sum_{j=1}^N(\delta_{r_j,I}-\delta_{r'_j,I})\nonumber\\
&&\times {\mathbf
  a}^*(K^{(N)}) \widehat{w}_N(-K^{(N)},L^{(N)}) {\mathbf
  a}(L^{(N)}),
\label{eq:5.8}
\end{eqnarray}
where $\delta$ is the Kronecker symbol and
$L^{(N)}=(L_1,\ldots,L_N)$, $L_j=(\ul_j,r'_j)$. 
Using that $\omega^0$ is invariant under $A\mapsto e^{isN^I}A e^{-isN^I}$, we
find that 
\begin{equation}
\omega^0([N\otimes\id, W_l])=0,\ \ \ \mbox{for all $l$.}
\label{eq:5.9}
\end{equation}
Thus ${\cal J}_{1,1}={\cal J}_{1,2}=0$.  
Next, we calculate ${\cal J}_{2,2}$. Recalling that $W_1(t)=\alpha_{-t}^0(W)$ and equation (\ref{eq:5.3}), we write 
\begin{eqnarray*}
\lefteqn{ 
[W_1(t), [N\otimes\id, W_1]] =\sum_{r, r', s, s'=
    I,II}(\delta_{s,I}-\delta_{s',I}) 
\int_{{\mathbb R}^{4d}}d\uk\  d\ul\  d\uk'\  d\ul'\ e^{-i(|\uk|^2-|\ul|^2)t}} \\
&&\times \widehat{w}_1((-\uk,r),(\ul,r'))\ \ \widehat{w}_1((-\uk',s), (\ul',s'))
\ \  [a^*(\uk,r)a(\ul,r'), a^*(\uk',s)a(\ul',s')]. 
\end{eqnarray*}
We expand the commutator on the right side and apply the state $\omega^0$ to obtain 
\begin{eqnarray*}
\lefteqn{
{\cal J}_{2,2}=\int_0^\infty dt \int_{{\mathbb R}^{2d}} d\uk\ d\ul \
\widehat{w}_1((-\uk,II), (\ul,I))\ \widehat{w}_1( (-\ul,I), (\uk,II))}\nonumber\\
&&\ \ \ \ \ \ \times\left\{ e^{-i(|\uk|^2-|\ul|^2)t} +
  e^{i(|\uk|^2-|\ul|^2)t}\right\}\left(\rho_{II}(\uk)-\rho_I(\ul)\right). 
\end{eqnarray*}
Because $W_1$ is selfadjoint, we have the relation
\begin{equation*}
\overline{\widehat{w}_1((-\uk,II),(\ul,I))}=\widehat{w}_1((-\ul,I), (\uk,II));
\end{equation*}
using this and the formula $\int_{-\infty}^\infty dt\  e^{i\tau
  t}=2\pi\delta(\tau)$, one sees that (\ref{eq:5.10}) holds. 
It is useful to keep in mind that
\begin{equation*}
\rho_{II}(\uk)-\rho_I(\uk)=\frac{e^{\beta^I(|\uk|^2-\mu^I)}-e^{\beta^{II}(|\uk|^2-\mu^{II})}}{\left(
    e^{\beta^I(|\uk|^2-\mu^I)}+1\right)\left(
    e^{\beta^{II}(|\uk|^2-\mu^{II})}+1\right)}.
\end{equation*}

{\it The resistance. } Let $\beta^I=\beta^{II}$ and 
$\mu^I=\mu, \mu^{II}=\mu+\Delta\mu$, with $\Delta\mu$ 
small. We expand 
\begin{equation*}
\rho_{II}(\uk)-\rho_I(\uk)=\beta\ \Delta\mu \
e^{\beta(|\uk|^2-\mu)} (e^{\beta(|\uk|^2-\mu)}+1)^{-2} +O((\Delta\mu)^2).
\end{equation*} 
Retaining only the first order in $\Delta\mu$ in equation (\ref{eq:5.10})
gives  (\ref{eq:5.11}) and (\ref{resistance}). 
Let $T=1/\beta$ denote the temperature and assume that $\mu>0$. We see that
$R(\mu,\beta)\sim T$, as $T\rightarrow\infty$. Next, we  examine the
dependence of the resistance on $T$, for small $T$, in three dimensions,  
and where $\widehat{w}_1$ is a radial function in both variables
(i.e. $\widehat{w}_1$ depends only on $|\uk|$ and $|\ul|$). We then have 
\begin{equation*}
R(\mu,\beta)^{-1}=8\pi^3\beta\int_0^\infty dr\
  r |\widehat{w}_1((\sqrt{r},II), (\sqrt{r},I))|^2
  \frac{e^{\beta(r-\mu)}}{\left( e^{\beta(r-\mu)}+1\right)^2}.
\end{equation*}
The fraction in the integral equals
$-\beta^{-1}\partial_r(e^{\beta(r-\mu)}+1)^{-1}$ and it follows that
\begin{equation}
R(\mu,\beta)^{-1}=\int_0^\infty dr\ f'(r)
\frac{1}{e^{\beta(r-\mu)}+1},
\label{r}
\end{equation}
where $f'$ denotes the derivative of the function
\begin{equation*}
f(r):=8\pi^3 r |\widehat{w}_1((\sqrt{r},II), (\sqrt{r},I))|^2.
\end{equation*}
Let us split the integral in (\ref{r}) as
\begin{eqnarray}
R(\mu,\beta)^{-1}&=&\int_0^\mu f'(r)\frac{1}{e^{\beta(r-\mu)}+1}
+\int_\mu^\infty f'(r)\frac{1}{e^{\beta(r-\mu)}+1}\nonumber\\
&=&f(\mu)-\int_0^\mu \ f'(r)\frac{1}{1+e^{-\beta(r-\mu)}}
+\int_\mu^\infty f'(r)\frac{1}{e^{\beta(r-\mu)}+1}.\nonumber\\
\label{r2}
\end{eqnarray}
Apply the change of variables $t=-\beta(r-\mu)$ and $t=\beta(r-\mu)$ in
the first and second integral on the right side of (\ref{r2}),
repectively. Then one has
\begin{eqnarray*}
\lefteqn{
R(\mu,\beta)^{-1}=f(\mu)}\\
&&+\frac{1}{\beta} \int_0^\infty dt\ \left\{
  f'(t/\beta+\mu)-f'(-t/\beta+\mu)\chi(t\leq\beta\mu)\right\}(e^t+1)^{-1}
\end{eqnarray*}
and using the mean value theorem, 
\begin{equation*}
R(\mu,\beta)^{-1}=f(\mu)+\frac{2}{\beta^2}\left(\int_0^\infty dt\
 f''(\xi_t)\frac{t}{e^t+1} +O(e^{-\beta\mu})\right),
\end{equation*}
for some $\xi_t\in[-t/\beta+\mu,t/\beta+\mu]$ and where the exponentially
small remainder term comes from removing the cutoff function
$\chi(t\leq\beta\mu)$. Retaining the main term ($\beta\rightarrow\infty$)
yields 
\begin{equation*}
R(\mu,\beta)^{-1}\approx f(\mu)+\frac{\pi^2}{6\beta^2} f''(\mu)
\end{equation*}
and consequently,
\begin{equation*}
R(\mu,\beta)\approx\frac{1}{f(\mu)+\pi^2 T^2 f''(\mu)/6},\ \ \ T\rightarrow 0.
\end{equation*}
At zero temperature, the resistance has the value $R(\mu,\infty)=f(\mu)^{-1}$
and it increases or decreases with increasing $T$ according to whether
$f''(\mu)<0$ or $f''(\mu)>0$.\\

{\bf Energy current.\ } The energy current
\begin{equation*}
{\cal P}=\omega_{\rm stat}(-i[H\otimes\id, W])=\omega^0(\sigma_+(-i[H\otimes
\id, W]))
\end{equation*}
has been introduced in equations (\ref{eq:2.91}), (\ref{eq:2.98}), see also
(\ref{eq:2.88}). We set $\hbar=1$.  Using 
the CAR and expression (\ref{eq:5.4}) for
$H$, one obtains
\begin{eqnarray*}
[H\otimes\id, W_N]&=&\int d K^{(N)}
d L^{(N)}\sum_{j=1}^N(|\uk_j|^2\delta_{r_j,I}-|\ul_j|^2\delta_{r'_j,I})\\
&& \ \times {\mathbf
  a}^*(K^{(N)}) \widehat{w}_N(-K^{(N)},L^{(N)}) {\mathbf
  a}(L^{(N)}),
\end{eqnarray*}
and it is readily verified that ${\cal P}_{1,1}={\cal P}_{1,2}=0$, and a
similar calculation as for the particle current shows that
\begin{equation}
{\cal P}_{2,2} =\int_0^\infty dt\
\omega^0([W_1(t),[H\otimes\id, W_1]])
\label{eq:5.12}
\end{equation}
is given by (\ref{eq:5.14}).\\


\end{document}